\documentclass[conference]{IEEEtran}

\usepackage{cite}
\usepackage{graphicx}
\usepackage{amssymb}
\usepackage{amsmath}
\usepackage{algorithmic}
\usepackage{algorithm}

\usepackage{array}
\usepackage{url}

\usepackage{lipsum}

\usepackage{tikz}
\usetikzlibrary{arrows.meta}
\usetikzlibrary{positioning}

\newtheorem{problem}{Problem}

\usepackage[caption=false,font=footnotesize]{subfig}

\usepackage[utf8]{inputenc}

\begin{document}
\title{Fast failover of multicast sessions in software-defined networks}

\author{\IEEEauthorblockN{Jorik Oostenbrink}
\IEEEauthorblockA{Delft University of Technology\\
Email: J.Oostenbrink@student.tudelft.nl}
\and
\IEEEauthorblockN{Niels van Adrichem}
\IEEEauthorblockA{Delft University of Technology\\
Email: N.L.M.vanAdrichem@tudelft.nl}
\and
\IEEEauthorblockN{Fernando Kuipers}
\IEEEauthorblockA{Delft University of Technology\\
Email: F.A.Kuipers@tudelft.nl}}

\maketitle

\begin{abstract}
With the rapid growth of services that stream to groups of users comes an increased importance of and demand for reliable multicast. In this paper, we turn to software-defined networking and develop a novel general-purpose multi-failure protection algorithm to provide quick failure recovery, via Fast Failover (FF) groups, for dynamic multicast groups. This extends previous research, which either could not realize fast failover, worked only for single link failures, or was only applicable to static multicast groups. However, while FF is know to be fast, it requires pre-installing back-up rules. These additional memory requirements, which in a multicast setting are even more pronounced than for unicast, are often mentioned as a big disadvantage of using FF. 

We develop an OpenFlow application for resilient multicast, with which we study FF resource usage, in an attempt to better understand the trade-off between recovery time and resource usage. Our tests on a realistic network suggest that using FF groups can reduce the recovery time of the network significantly compared to other methods, especially when the latency between the controller and the switches is relatively large. 
\end{abstract}

\section{Introduction}
\IEEEPARstart{S}{ervices} like IPTV, which have to stream to many subscribers at the same time, have been gaining in popularity and could benefit a great deal from multicast \cite{cha2006case}. For example, Twitch, an online service allowing individuals to stream to a group of viewers, had a peak concurrent viewer count of more than 2 million in 2015 \cite{TwitchViewers2015}. 

Unfortunately, mostly because of its decentralized nature, traditional IP multicast is lacking in reliability, scalability and security \cite{diot2000deployment, li2013scaling ,iyer2014avalanche}. 
The recent emergence of Software-Defined Networking (SDN) provides a new opportunity to implement reliable, scalable and secure multicast support. In SDN the control plane, which decides where traffic is sent, is logically centralized and decoupled from the data plane, which forwards the data. An SDN controller collects information from and manipulates traffic forwarding settings of all networking devices in a network, and thus makes its decisions based on a centralized view of the network.

This allows it to overcome most of the problems traditional IP multicast experiences. Switches no longer have to communicate with each other to distributively form a multicast tree. With its overarching view, the controller can potentially compute a much more efficient tree. The controller can also perform admission control, as it can decide which hosts to allow to join a group. In addition, it can reroute traffic to always pass through certain devices if necessary. The advantages of using SDN for multicast applications have been shown in a variety of circumstances \cite{iyer2014avalanche, bondan2013multiflow, arefin2013opensession, zhao2014software, noghani2014streaming, egilmez2012openqos}.

In this paper, we discuss the use of Fast Failover (FF) groups to provide fault tolerance against multiple ($F$) link failures for dynamic multicast groups. Fast Failover has been extensively researched for unicast purposes, e.g. see \cite{van2014fast}, but not much research has been done on using FF in a multicast setting. In addition, providing fault tolerance support for dynamic groups, where subscribers can join and leave at any time, brings its own unique challenges.

In Fast Failover a switch can rapidly switch over to a backup port after detecting link failure, without any input from the controller. This allows for very rapid failure recovery, but unfortunately also takes up a lot of resources/memory, as backup paths need to be calculated in advance and the corresponding backup rules need to be stored in the switches \cite{Niels2}. Although extensive studies to the precise resource consumption are missing, this additional memory overhead is frequently listed as the main reason not to use FF for multicast traffic. We will investigate the resource usage of FF on a real-world topology, to better understand the trade-off between resource usage and recovery speed. In order to do so, we will implement our own multicast controller in OpenFlow (OF), which is a popular protocol enabling the communication of a controller with the data plane (i.e., the OpenFlow switches) \cite{OpenFlowSite, mckeown2008openflow}. Through OpenFlow, a controller can install flow entries to direct the flow of traffic. When a packet gets matched to a flow entry, the instructions corresponding to this entry get executed. These instructions can for example be to output the packet to a specific port or to add a VLAN tag.

Our main contributions in this paper are organized as follows: Section \ref{related} discusses related work on fault tolerance in SDN. Our problem definition is presented in Section \ref{approach}. In Section \ref{GPA_section}, we propose a general-purpose algorithm to provide Fast Failover support for dynamic multicast trees and compare it to existing solutions. In Section \ref{Implementation}, we incorporate this algorithm as part of an OpenFlow controller application providing $F$-link fault tolerant multicast support and, in Section \ref{Evaluation}, we evaluate the performance of this implementation in terms of resource usage and recovery time. We conclude in Section \ref{Conclusions}.

\section{Related Work}
\label{related}
Fault tolerance and recovery are important elements of any networking technology. As a consequence, also a lot of research has been done on fault tolerance in SDN, e.g. see \cite{recodis}. However, most of this research has focused on unicast and not multicast messaging. 

In general, recovery techniques can be divided into two categories: restoration and protection. Restoration is almost completely reactive, new paths are configured only after a link or node failure. Protection is a more proactive approach, where backup paths are installed in advance and can be quickly switched to after failure. For an overview of recovery algorithms we refer to \cite{kuiperssurvey}.

\subsection{Unicast Failure Recovery in SDN}
%unicast restoration
Some of the earliest work in OpenFlow fault recovery was done by Dimitri Staessens et al. \cite{staessens2011software}. They evaluated the time it takes for the controller to restore a network from failure. In their experiments this took between 260 and 310 ms. Of this time around 170 ms was the time it took for the controller to detect the failure. The recovery procedures themselves were completed in 80-130 ms. Their conclusion was that, because of OpenFlow's dependency on a centralized controller, it will be hard to achieve a 50 ms restoration (the carrier grade goal) on large networks serving many flows. In order to achieve a sub 50 ms recovery time on such networks, protection is required.

The same authors later compared restoration and protection on emulated networks running on physical Linux nodes \cite{sharma2013openflow}. Their protection method, using BFD to detect failures and a Fast Failover mechanism in the switches to quickly switch to the backup paths, achieved a recovery time below 50 ms. Most of this time was taken up by the BFD failure detection time of 40 to 44 ms.

Maciej Ku\'{z}niar et al. proposed a somewhat different approach to failure recovery \cite{kuzniar2013automatic}. Upon detecting failure, their runtime system spawns a new controller in an emulated environment consisting of the complete network topology minus the failed elements. By replaying all previous inputs to the controller, this emulated network is configured into a state that accounts for the failed elements. The real network can now be recovered by installing the difference between the rulesets of the real and emulated network and replacing the old controller with the new one. This approach allows network developers to write failure-agnostic code.

The average automatic failover time of an OpenFlow switch can be as low as 19.63 ms, but also as high as 1564.33 ms, depending on the model of the switch and the type of link \cite{lee2014path}. This delay is mostly caused by the failure detection time. To decrease the detection time, Steven S. W. Lee et al. implemented a monitoring system in which the controller probes the network to detect failures \cite{lee2014path}. The controller continuously sends packets over monitoring paths in the network. If these packets do not reach the controller, it tries to find the specific failure point by sending FLI packets through the affected path. Every switch does not only forward these FLI packets, but also copies and sends them to the controller. This way, after sending multiple FLI packets to prevent false positives, the controller knows exactly which link is down and the network can be recovered. To prevent monitoring packets from being lost due to congestion and to reduce their jitter, they are sent through a QoS queue.

%unicast protection
James Kempf et al. proposed extending OpenFlow switches with end-to-end failure discovery functionality \cite{kempf2012scalable} to enable better decentralized protection schemes. By using an MPLS approach, they managed to get a failover time of around 28.2 ms.

Andrea Sgambelluri et al. also implemented a protection scheme based on their own proposed extension to OpenFlow \cite{sgambelluri2013openflow}. They proposed a feature which enables OpenFlow switches to automatically remove and reject any flows with a faulty link or output port. This way, backup paths can be installed by adding backup flows with a lower priority than the primary flows to the switches. When a link or port fails, the switch will automatically remove any affected flows and thus use the backup flows for new incoming packets. This approach was evaluated on both Mininet and a real network. The average recovery time was 79.443 ms on Mininet, and 32.74 ms on the real network, but was only determined by the failure detection time.

In \cite{capone2015detour}, the authors proposed using the OpenState extension to OpenFlow instead of Fast Failover groups. If, after a failure, a packet gets sent back upstream over the primary path, this allows the switches the packet travels back through to immediately send all subsequent packets over the backup path, instead of first sending them towards the faulty link. Normally, it is not possible to let OpenFlow switches communicate with each other like that. 

FatTire (Fault Tolerating Regular Expressions) \cite{reitblatt2013fattire} is a new language specifically designed for writing fault-tolerant OpenFlow network programs. It allows programmers to specify legal paths through the network, along with fault-tolerance requirements for those paths. Backup paths are computed and pre-installed per link as required. Switches automatically switch from the primary to the backup paths when a link is down by using the Fast Failover functionality of OpenFlow. In an experiment in Mininet, the transfer completion time after fast failover was only slightly higher than in the case of no failure.

Niels L. M. van Adrichem et al. proposed a fast failover scheme using per-link BFD for failure detection for single link fault tolerance \cite{van2014fast}. They noted a few advantages of this approach: per-link monitoring minimizes the round-trip time, thus allowing for a lower worst-case failure detection time. Also, a BFD control packet plus encapsulation consists of at most 720 bits and a single link can only have one BFD session (as a per-link approach is used), thus the overhead of this method is quite small. On a testbed consisting of software switches their implementation achieved a maximal recovery time of 6.7 ms on a simple topology and a maximal recovery time of 4.8 ms on a (larger) ring topology. As per-link sessions are used instead of per-path sessions, the recovery time does not increase when the network grows or the path length increases. Their work was continued in \cite{Niels2} where all-to-all path computing algorithms were given that account for (single) link or node failure, as well as an algorithm which accounts for both.

A different approach to protection was given in \cite{kitsuwan2015independent} by Nattapong Kitsuwan et al. Their method uses two ``planes:'' a working plane and a transient plane. The working plane is used for routing when there are no failures. If a link fails, all packets that would be sent over this link now get transported over the transient plane till the network has been restored. Tags (VLAN or MPLS) are added to the packet header to indicate which link has failed. This switch from working to transient is done on command by the controller, not by using a Fast Failover group. The transient plane itself contains the shortest paths to all possible destinations for all possible single link failures and is computed and installed beforehand. In their experiments on a pan-European network there was no packet loss and the switchover time was around 23.5 ms. 

Michael Borokhovich et al., in \cite{borokhovich2014provable}, presented multiple failover implementations for OpenFlow that achieve ``maximum robustness'' by treating the problem of forwarding a packet to its destination in an unknown graph as a graph searching problem. These algorithms have the interesting property that ``connectivity is preserved under arbitrary link failures, subject to the (weakest and necessary) condition that the underlying physical network be connected.'' However, in larger networks, the number of tags (as well as the number of flows) required by their algorithms becomes very large. In addition, the path that packets take to reach their destination after a link failure can be substantially longer than optimal.

\subsection{Multicast Failure Recovery in SDN}
In \cite{gyllstrom2014recovery} a set of algorithms was proposed to make multicast trees robust to single link failures. Failed links are detected by the controller by periodically checking if the rate of packet loss exceeds a certain threshold. Backup trees are calculated to maximize the reuse of links of the primary tree, subject to some constraints. As this is an NP-hard problem, an approximation algorithm was used. 
Both a restoration and a protection algorithm were proposed. The protection algorithm pre-installs backup flows, but does not use Fast Failover groups. Instead, the controller changes a single flow in the root nodes of the primary tree to switch over to a new tree. Finally, an algorithm was proposed to merge flows in the same switch, where possible, to further reduce the amount of flow entries required. While their merging algorithm is effective for their reactive approach, their failure detection algorithm takes up a large amount of processing time in the switches, which limits the size of the network that can be monitored.

Daisuke Kotani et al. also proposed pre-installing a primary tree and multiple backup trees in the network and switching (by changing a flow in the root nodes) when necessary \cite{kotani2012design, DaisukeKotani2016}. The trees can be differentiated from each other by changing the packet headers. 
In contrast to \cite{gyllstrom2014recovery}, their method does allow adding and removing subscribers dynamically. To decrease the time it takes to connect a new subscriber to a tree, primary and backup trees connected to all possible recipients are precomputed. When a new node needs to be added to the tree, the necessary flows can be installed immediately. Their results on a virtual network were that precomputing gives some overhead at the beginning, but decreases the computing time for adding other receivers significantly. 
They evaluated their implementation on a network consisting of one physical switch and a few software switches. The average maximum packet loss duration was $\leq$ 10 ms for 4 and around 40 ms for 10 multicast groups. However, in their experiments the latency between the controller and the switches was very low, which, as we will indicate, influences the results. 

The approach taken by Vignesh Renganathan Raja et al. is similar \cite{subtree}. They proposed dividing the multicast tree into subtrees and installing backup paths for these subtrees instead of for the complete tree. Also, instead of computing disjoint trees, they compute multiple disjoint paths for each leaf of the subtree. If a link or node fails, the controller first searches for the subtree it needs to recover, and then modifies the necessary flow tables to switch over to the appropriate paths. 
The authors evaluated their approach via Mininet, but ignored failure detection time and had zero latency between the controller and the switches.

The approach most similar to ours was taken by Thomas Pfeiffenberger et al. in \cite{pfeiffenberger2015reliable}. Per link in the multicast tree, a backup tree is computed and installed. VLAN tags are used to differentiate different trees and Fast Failover groups are used to quickly switch over to the backup trees after a link failure. However, their approach 
assumes that all subscribers are known in advance and it cannot dynamically add and remove subscribers; a functionality we provide in this paper. In addition, we provide a more extensive look at the resource usage of Fast Failover in both single-link and three-link fault tolerance cases.

\section{Problem Definition}
\label{approach}

As becomes apparent from the related work, a protection scheme, e.g. via Fast Failover, leads to faster recovery times than a restoration scheme (which first has to ask the controller how to react).

Fast Failover in OpenFlow is enabled by installing FF groups in the switches. A FF group contains a list of action buckets, each of these buckets is associated with a port and/or another FF group. A bucket is considered live if either its port or its group is live and a group is considered live if at least one of its buckets is live. Applying a Fast Failover group is equivalent to applying the actions of its first live action bucket. 

In this paper, we will assume link monitoring with BFD is used as the liveness mechanism, as it has been shown that this enables very low worst-case failure detection times at the cost of not much overhead \cite{van2014fast}. 

Switches cannot inform each other of link failures, so packets must be rerouted from the point of failure until the controller has restored all paths. We define this problem formally as follows: 

\begin{problem}
Given a network $G = (V,E)$, where $V$ is the set of nodes and $E$ the set of links, a root $r \in V$, a fault tolerance requirement $F \in \mathbb{N}$, and a sequence of multicast join/leave requests $r_0 = \emptyset, r_1, r_2, \dots \subseteq V$, where $|r_i \setminus r_{i+1}| + |r_{i+1} \setminus r_i| = 1$ for all $i$. Compute a sequence of primary and backup flow entries $f_i$ and FF groups $g_i$, without knowledge of any future requests $r_j \forall j > i$, such that all nodes $r_i$ are connected to $r$ by $f_i$ and $g_i$ under all combinations of $F$ or less link failures.
\end{problem}

Even if we are able to minimize the amount of flow entries needed, to get an $F$-link fault-tolerant network, we must account for all possible combinations of $F$ failing links. This takes a large toll on resource usage, not only in the form of TCAM memory in the switches, but also in the form of the computation and installation time of all trees.

In the next section, we will present a general-purpose algorithm to compute these trees and which can be combined with many existing tree construction algorithms.

\section{General-Purpose Algorithm}
\label{GPA_section}
Different networks and applications have widely different requirements on their multicast trees. In some cases simple shortest path trees suffice, while in other cases the tree construction algorithm has to account for bandwidth constraints and security issues. To support fault tolerance, instead of having to completely reinvent the wheel and come up with a new tree construction algorithm for every specific case, it would be better to have a general-purpose fault-tolerance algorithm that can be combined with already existing tree construction algorithms.

In this section, we propose an algorithm to provide $F$-link fault tolerance that can be combined with any existing tree algorithm that works on a request-by-request basis and does not rearrange the tree. With which we mean that links are only added after join requests and are only removed after leave requests. 

$F$ can be varied per multicast tree, so the amount of fault tolerance necessary can be decided on a group-by-group basis. 

We assume that the tree construction algorithms resolve a Join function as described in Algorithm \ref{Join}.  
\begin{algorithm}[h]
\caption{Join}
\label{Join}
\begin{algorithmic}
\REQUIRE $G = (V,E)$, $T = (root,V^T,E^T)$, $v \in V$
\STATE Where $G$ is the network, $T$ is a tree with root $root$ in the network and $v$ is a host in the network.
\ENSURE $path$
\STATE Where $path$ is the path from $root$ to $v$ in $T$ if $v$ were to be added to tree $T$. If $v$ can't (or isn't allowed to) join the group, $\emptyset$ should be returned instead.
\end{algorithmic}
\end{algorithm}

Note that this function does not actually add $v$ to the tree, but just returns the path required to do so. As there are no rearrangements, this path contains all the information needed to properly do this. 

\subsection{Tree implementations}
To showcase and evaluate our algorithm, we will implement the join function for two different trees: the Shortest Paths Tree (SPT) and an approximation tree for the Dynamic Steiner Tree (DST) problem \cite{doi:10.1137/0404033}. 

\subsubsection{Shortest Paths Tree}
Given a network $G$. A shortest paths tree with root $root$ is a tree $T$ such that the distance in $T$ from $root$ to any other vertex $v \in T$ is a shortest path from $root$ to $v$ in $G$.

An important implementation detail of a shortest paths tree is the choice of link cost. The simplest decision is to set every cost to $1$, thus computing the minimum hop paths from root to destinations. However, its also possible to use other costs, like the traffic load on or the latency of a link.

In \cite{noghani2014streaming} it was shown that the packet loss under network congestion of an SPT using the traffic load on links as costs is very similar to that of an SPT using the amount of hops as distance. We chose to implement the minimum hop variant, see Algorithm \ref{JoinSPT}. 

\begin{algorithm}[h]
\caption{SPT Join}
\label{JoinSPT}
\begin{algorithmic}
\IF{$v \in V^T$}
\RETURN $\emptyset$
\ENDIF
\STATE $\epsilon \leftarrow \frac{1}{|E^T| + 1}$
\STATE Define $c : E \to \mathbb{R}$
\STATE Where $c(e) = \begin{cases}1 - \epsilon & \text{if } e \in E^T\\1 & \text{otherwise}\end{cases}$
\STATE $path \leftarrow$ Dijkstra($G$, $root$, $v$, $c$) \COMMENT{Use Dijkstra's algorithm to compute the shortest path from $root$ to $v$ in $G$ given link costs $c$}
\IF{a path does not exist}
\RETURN $\emptyset$
\ENDIF
\RETURN $path$
\end{algorithmic}
\end{algorithm}

Note that while we use $1$ as the cost of most links in our Dijkstra computation, we use $1 - \epsilon$ as cost for the links that are already in the tree. This is done to ensure that the path returned by Dijkstra re-uses as much of the links already in the tree as possible. As long as $\epsilon < \frac{1}{|E^T|}$, this still results in a minimum hop path.

\begin{IEEEproof}
Assume we have a shortest path $p$ of length $L$, computed with the cost function as described above, where $\epsilon < \frac{1}{|E^T|}$. Assume $p_m$ is a minimum hop path with length $L_m$. We need to prove that $L = L_m$.

Let $c(p)$ and $c(p_m)$ indicate the cost of $p$ and $p_m$ as computed by the cost function and let $c(l)$ be the cost of link $l$. We know $c(p) \leq c(p_m)$. 

Assume $L_m < L$, then
\begin{equation*}
\begin{split}
c(p) & = \sum\limits_{l \in p}c(l) \\
& = L - \epsilon|\{l \in E^T | l \in p\}|\\
& > L - \frac{1}{|E^T|}|E^T|\\
& = L-1 \\
& \geq L_m\\
& \geq L_m - \epsilon|\{l \in E^T | l \in p_m\}|\\
& = c(p_m)
\end{split}
\end{equation*}
This is in contradiction with $c(p) \leq c(p_m)$, so $L = L_m$ and $p$ is a minimum hop path.
\end{IEEEproof}

The time complexity of the join function is the same as that of Dijkstra's algorithm: $O(|E| + |V| \log |V|)$.

\subsubsection{Dynamic Steiner Tree}
The Dynamic Steiner Tree (DST) problem, proposed by Makoto Imase et al. in \cite{doi:10.1137/0404033}, is the problem of, given a sequence of join and leave requests, finding a minimum-cost tree after each request connecting the terminal set and without knowledge of any future requests.

They proposed a greedy algorithm \cite{doi:10.1137/0404033}, see Algorithm \ref{JoinDST}, for approximating a solution to the DST problem. The idea behind this algorithm is simple: join new nodes by the shortest path to a nearest node already in the tree.

We choose to use $1$ as the cost of every single link, thus the algorithm tries to minimize the amount of links in the tree. 
By minimizing the amount of links in our trees we do not only minimize the amount of memory taken up by the tree itself, but we also minimize the amount of backup paths that have to be calculated and installed in the network. 

\begin{algorithm}[h]
\caption{DST Join}
\label{JoinDST}
\begin{algorithmic}
\IF{$v \in V^T$}
\RETURN $\emptyset$
\ENDIF
\STATE $w \leftarrow$ closest node to $v$ in $T$
\COMMENT{As determined by Dijkstra's algorithm}
\STATE $path \leftarrow$ shortest path from $w$ to $v$ in $G$
\COMMENT{Also calculated by Dijkstra's algorithm}
\IF{a path does not exist}
\RETURN $\emptyset$
\ENDIF
\STATE $pre \leftarrow$ path from $root$ to $w$ in $T$
\RETURN $pre + path$
\end{algorithmic}
\end{algorithm}

The time complexity of this function is $O(|E| + |V| \log |V|)$

\subsection{Algorithm}
Since we update the primary and backup trees on a request-by-request basis, at any given time we only have to compute the necessary flows to add one subscriber to the multicast group. This 
reduces the problem complexity from finding complete primary and backup trees to finding primary and backup paths.

To ensure the delay between subscribing to a group and receiving the first packet is as low as possible, the primary path is computed and installed first.

To reach $F$-link protection, our general-purpose algorithm, see Algorithm \ref{Alg1}, first calculates and installs backup paths for all links in the primary path. Then, to protect every link on these backup paths, we also compute backup paths for those links, etc. This goes on till the required fault tolerance level $F$ has been reached (if possible). In practice, this means that backup trees are installed for every single link in a protected tree. When protecting a path to a new subscriber, backup trees for links that already are protected are updated, while new backup trees, with a single subscriber, are constructed for unprotected links of the path. 

\begin{algorithm}[h]
\caption{Primary and Backup Tree Computations}
\label{Alg1}
\begin{algorithmic}
\REQUIRE $T = (root, V^T, E^T)$, $v$, $r$, $F$
\STATE Where $v$ is a host, $r$ is either ``join'' or ``leave'' to indicate if $v$ wants to join or leave $T$ and $F$ is the required amount of link fault tolerance
\IF{$r$ = ``leave''}
\STATE Leave($T$, $v$)
\RETURN
\ENDIF
\STATE $path \leftarrow$ Join($G$, $T$, $v$)
\IF{$path = \emptyset$}
\RETURN
\ENDIF
\STATE InstallFlows($path$)
\STATE $T \leftarrow T \cup path$
\STATE $queue \leftarrow$ new Queue()
\IF{$F > 0$}
\STATE $queue$.Enqueue($(path, T, \emptyset)$)
\ENDIF
\WHILE{$queue$ is not empty}
\STATE $(path, tree, down) \leftarrow$ $queue$.Dequeue()
\FORALL{$e = (x,y) \in path$}
\IF{$e$ in $tree$ does not have a backup tree}
\STATE $tree.e.backup \leftarrow (x, \{x\}, \emptyset)$
\STATE $tree.e.backup.tag \leftarrow T.nextTag$
\STATE $T.nextTag \leftarrow T.nextTag + 1$
\ENDIF
\STATE $L \leftarrow down \cup \{e\}$
\STATE $b\_path \leftarrow $ Join($G - L$, $tree.e.backup$, $v$)
\IF{$b\_path \neq \emptyset$}
\STATE InstallFlows($b\_path$)
\STATE $tree.e.backup \leftarrow tree.e.backup \cup b\_path$
\IF{$|L| < F$}
\STATE $queue$.Enqueue($(b\_path, tree.e.backup, L)$)
\ENDIF
\ENDIF
\ENDFOR
\ENDWHILE
\end{algorithmic}
\end{algorithm}

The probability that a link fails is relatively low, so computing and installing all backup paths is allowed to take longer than computing and installing the primary path. The fault tolerance of the multicast group (with respect to $v$) is built up slowly. First all backup trees for the primary tree are updated, then the backup trees for these trees are updated, etc.

As the same algorithm is used both to compute the primary and the backup trees, the trees satisfy exactly the same constraints.

To differentiate different trees (with the same primary tree) from each other in the flow tables, each backup tree gets a unique VLAN tag.

InstallFlows installs all flows from $path$ that are not yet installed. This could be combined with adding the path to the tree.

Removing a subscriber $v$ from a tree is the same in every single tree construction algorithm without rearrangements (except those that leave some unnecessary links remaining): Remove $v$ itself plus all links and nodes that lead to $v$, but not to any other terminal node.

To prevent any unnecessary packets being sent over the network, the Leave function described in Algorithm \ref{Leave} first removes the necessary primary flows, followed by all backup flows.

\begin{algorithm}[H]
\caption{Leave}
\label{Leave}
\begin{algorithmic}
\REQUIRE $T$, $v$
\IF{$v \notin V^T$ or $v = root$}
\RETURN
\ENDIF
\STATE $cur \leftarrow v$
\WHILE{outdegree($cur$) $\leq 1$ and $cur \neq root$}
\STATE $pre \leftarrow$ predecessor($cur$)
\STATE RemoveFlow($(pre, cur)$)
\STATE $cur \leftarrow pre$
\ENDWHILE
\STATE $cur \leftarrow v$
\WHILE{$cur \neq root$}
\STATE $pre \leftarrow$ predecessor($cur$)
\STATE Leave($(pre, cur).backup$, $v$) 
\STATE $cur \leftarrow pre$
\ENDWHILE
\STATE $cur \leftarrow v$
\WHILE{outdegree($cur$) == 0 and $cur \neq root$}
\STATE $pre \leftarrow$ predecessor($cur$)
\STATE $T \leftarrow T - cur$
\STATE $cur \leftarrow pre$
\ENDWHILE
\end{algorithmic}
\end{algorithm}

It takes only one join to compute the primary path, so in our case the primary path is computed in $O(|E| + |V| \log |V|)$ time. However, it takes at most $O(|E|^F)$ joins to compute all the backup paths. This means that it not only takes a long time to compute all backup paths, but that this method also requires a lot of memory in the switches to store all flow entries and FF groups and a large tag space to differentiate all trees. Fortunately, $F$ would generally be set low, as it is expected that not many links would fail at the same time.

\subsection{Comparison with Other Methods}
\begin{table*}[!t]
\renewcommand{\arraystretch}{1.3}
\caption{Comparison of algorithm \ref{Alg1} to other recovery methods}
\label{table_comparison}
\centering
\begin{tabular}{c|c|c|c|c}
 & \bfseries Restoration & \bfseries Fast Tree Switching \cite{kotani2012design, DaisukeKotani2016} & \bfseries Fast Failover (Algorithm \ref{Alg1})\\ 
\hline \hline
\bfseries Resource usage & + + & + & - - \\
\hline
 Computation time & + + & + & - - \\
 Flow entries & + + & + & - - \\
 Group tables & + + & + + & - - \\
 Tags & + + & + & - - \\
\hline
\bfseries Recovery Time & - - & - & + + \\
\hline
 Affected by RTT controller and switch & - - & - - & + + \\
Affected by number of multicast groups & - - & - & + + \\
\end{tabular}
\end{table*}

In Table \ref{table_comparison}, we compare our Algorithm \ref{Alg1} to two other known methods for multicast fault tolerance that support adding subscribers to a group on demand:
\begin{itemize}
\item Restoration: not pro-actively installing any backup paths, but installing and removing flow entries to restore the network after a failure.
\item Fast tree switching \cite{kotani2012design,DaisukeKotani2016}: a protection scheme in which multiple disjoint trees per group are installed in the network. When a link fails, the controller switches the group to a tree without any link failures by changing a flow in the root of the trees.
\end{itemize}

\subsubsection{Resource Usage}
Restoration requires the least resources of all methods, as no backup paths have to be pre-computed and pre-installed. Fast tree switching also does not require a lot of resources, as only $F+1$ (1 primary + $F$ backup) trees have to be installed. For the Fast Failover algorithm, when only single link fault tolerance is required the amount of resources used is quite small. However, when requiring more fault tolerance this very quickly increases and it may not be possible to support large multicast groups.

One of the most scarce resources in OpenFlow switches are group tables. Both restoration and fast Tree Switching do not require any Fast Failover groups, thus having an advantage in this category.

\subsubsection{Recovery Time}
One of the main reasons to use Fast Failover groups instead of other methods is to allow for a very small recovery time. By using FF groups, the network directs traffic to the backup trees almost as soon as the failure is detected by the switches. In contrast, when using the fast tree switching method the controller first has to receive a notification from the network that a link has failed, and then has to modify one flow entry per affected multicast group. As noted in Section \ref{related}, the recovery time when using restoration as a recovery method is very high. Not only does the controller first need to receive a notification about the failed link, it then has to compute and add/remove/modify all necessary flow entries to recover the network.

\section{Implementation}
\label{Implementation}
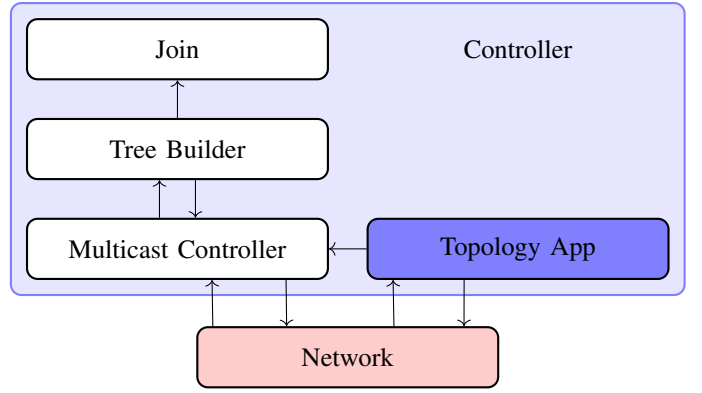
\begin{figure}[t!]
\centering
\begin{tikzpicture}
\tikzstyle{bigbox} = [draw=blue!50, thick, fill=blue!10, rounded corners, rectangle]
\tikzstyle{box} = [minimum size = 0.8cm,minimum width=4cm, rounded corners,rectangle, fill=white,draw=black]
\matrix[row sep=5mm, column sep=5mm, inner sep=2mm, bigbox] (matrix) {
\node[box] (join) {Join}; & \node {Controller};\\
\node[box] (builder) {Tree Builder};\\ 
\node[box] (controller) {Multicast Controller}; & \node[box,fill=blue!50] (topo) {Topology App};\\
};

\begin{scope}[every node/.style=box]
	\node[below = 0.4 of matrix,thick,fill=red!20] (network) {Network};
\end{scope}

\path[->, draw=black]
	(controller) edge[in = -120, out = 120, looseness=0] node {} (builder)
	(builder) edge node {} (join)
	(builder) edge[in = 60, out = -60, looseness=0] node {} (controller)
	(topo) edge node {} (controller)
	(network) edge[in = -42, out = 167, looseness=0] node {} (controller)
	(controller) edge[in = 153, out = -16, looseness=0] node {} (network)
	(network) edge[in = 194, out = 34, looseness=0] node {} (topo)
	(topo) edge[in = 15, out = -150, looseness=0] node {} (network)
	;
\end{tikzpicture}
\caption{Architecture}
\label{arch}
\end{figure}

In this section, we describe the implementation of an OpenFlow application providing $F$-link fault tolerance for multicast video streams using the methods described above. The application (which is made available via GitHub \cite{JorikApp}) has been implemented in Python, using the Ryu Framework \cite{Ryu} to communicate with the OpenFlow network, and the NetworkX library \cite{NetworkX} to store the network topology and calculate shortest paths.  

The application is divided into three modules, see Figure \ref{arch}:
\begin{enumerate}
\item Multicast Controller: The interface between the application and the network. Installs flows, adds and removes hosts to and from groups, etc.
\item Tree Builder: Computes and installs primary and backup trees.
\item Join: The basic tree construction algorithm used by Tree Builder to compute paths (e.g. the SPT or DST algorithm).
\end{enumerate}
Furthermore, a topology app built-in in Ryu is used to gather all necessary topology information about the network. The topology app uses LLDP packets to periodically probe the network. This app also notifies the Multicast Controller module when a link has failed.

As the pseudocode for Tree Builder and Join has already been presented, in the following we only describe the functionality and implementation of the Multicast Controller.

\subsection{Multicast Controller}
\begin{table*}[!t]
\renewcommand{\arraystretch}{1.3}
\caption{Example of a copied FF group with backup ports 7, 8 and 9}
\label{exampleFFgroup}
\centering
\begin{tabular}{||c|c||}
\bfseries Watch Port & \bfseries Actions\\ 
\hline
1 & output:1\\
2 & tag=1,output:2\\
7 & tag=2,output:7\\
10 & tag=3,output:10\\
\end{tabular}
\begin{tabular}{||c|c||}
\bfseries Watch Port & \bfseries Actions\\ 
\hline
1 & Drop\\
2 & Drop\\
8 & tag=2,output:8\\
11 & tag=4,output:11\\
\end{tabular}
\begin{tabular}{||c|c||}
\bfseries Watch Port & \bfseries Actions\\ 
\hline
1 & Drop\\
2 & Drop\\
9 & tag=2,output:9\\
12 & tag=5,output:12\\
\end{tabular}
\end{table*}

The Multicast Controller is responsible for installing flow and group entries, adding and removing hosts to and from groups and generally facilitating all communication with the OpenFlow switches.

Currently our implementation only supports IPv4 multicast traffic and assumes hosts are using IGMPv3 to signal their intent to join or leave multicast groups. However, support for more protocols can easily be added by making some slight changes to this module.

Note that in IGMPv3 a single multicast group (or multicast-address, as is more applicable in this case) can contain multiple sources. A host can subscribe to any subset of these sources. For our purposes, it is best to treat every source+address combination as a separate multicast group, as they each require their own tree.

OpenFlow switches are instructed to send all packets that cannot be matched to any of their flow entries to the controller. This way the Multicast Controller receives all IGMP packets, as well as IPv4 multicast packets from a source+group it has not yet installed flow entries for.

When receiving an IPv4 multicast packet from a source + multicast-address combination it has not yet seen before from switch $A$, the Multicast Controller stores this combination and informs the Tree Builder that a new multicast group needs to be created rooted at $A$. It then installs a single low-priority flow at switch $A$ which drops all packets from this multicast group pair, to prevent the switch from continuously sending packets to the controller until a host has joined this group.

By listening to (and removing from the network) the IGMP membership reports the module knows exactly when a host wants to join or leave a specific group. It then instructs the Tree Builder to add or remove the host. The Tree Builder in turn calculates the primary and backup flows necessary to do so and instructs the Multicast Controller to install them.

Every flow entry can be used to send packets to multiple ports at once (using the Apply Actions instruction). When backup needs to be provided for a flow, a separate Fast Failover group gets installed for each of these ports and the flow entry moves the packet to these group tables instead. Backup ports can then be installed by adding more buckets to these groups. Each bucket contains an output port, a watch port (same as the output port) and if necessary an additional action to add/change the VLAN tag.

If $F \geq 2$ multiple FF group tables may need to be installed for a single port, that is, if the root of the backup tree of a link has multiple outgoing links, which themselves also need to be protected. While it is possible to install a bucket instructing the switch to send a packet over multiple outgoing links, the problem lies in detecting and reacting to the failure of one of these links. Preferably, we would like to only switch over to the backup tree for those links that have actually failed, and keep sending the packet over the links which are not down. This requires us to install a separate FF group for each of them.

Unfortunately, most OpenFlow switches do not support chaining multiple groups together. Our solution to this is as follows: assume we have a FF group $g$ to which we have to add $N$ backup ports, as backup for port $p$. The module first makes $N-1$ copies of this group, with the slight change that, instead of outputting the packet, all buckets of these copies drop the packet. Then it adds one of the necessary backup port buckets to each of these copies. An example of the resulting group tables is given in Table \ref{exampleFFgroup}. Finally the flow entry of this FF group is modified to also send the packet through the copy FF groups. When $p$ is still live the copies have no effect, as they just instruct the switch to drop the packet. When $p$ is down (in addition to any previous ports in the groups), the packet is sent to all $N$ backup ports.

There is one final property of OpenFlow we have to account for: When an action list contains both (a) actions to output to ports and (b) to send a packet through group tables, only the group actions get executed. In the worst case, 3 separate action lists need to be applied to send the packet properly to all output ports:
\begin{enumerate}
\item An action list containing only group actions
\item An action list containing only output to port actions
\item An action list containing output to port actions, but also a VLAN tag pop action, to remove the VLAN tag before being send directly to a host. 
\end{enumerate}

OpenFlow has the capability to divide flow entries across different flow tables. If a packet first arrives at the switch, it will only be matched to flow entries in the first table. Optionally, a flow entry in this table can instruct the switch to go to another table, which will result in the packet being matched to the entries in that table. To make sure the action lists are applied separately, Multicast Controller installs flow entries in up to 3 different flow tables. In each one of these flow tables, the switch gets instructed to apply an action list and, if necessary, to go to the next table.

\section{Evaluation}
\label{Evaluation}
In this section, we evaluate our algorithm on both resource usage and recovery time. 

\subsection{Setup}
\begin{figure}[t!]
\centering
\includegraphics[width=\linewidth]{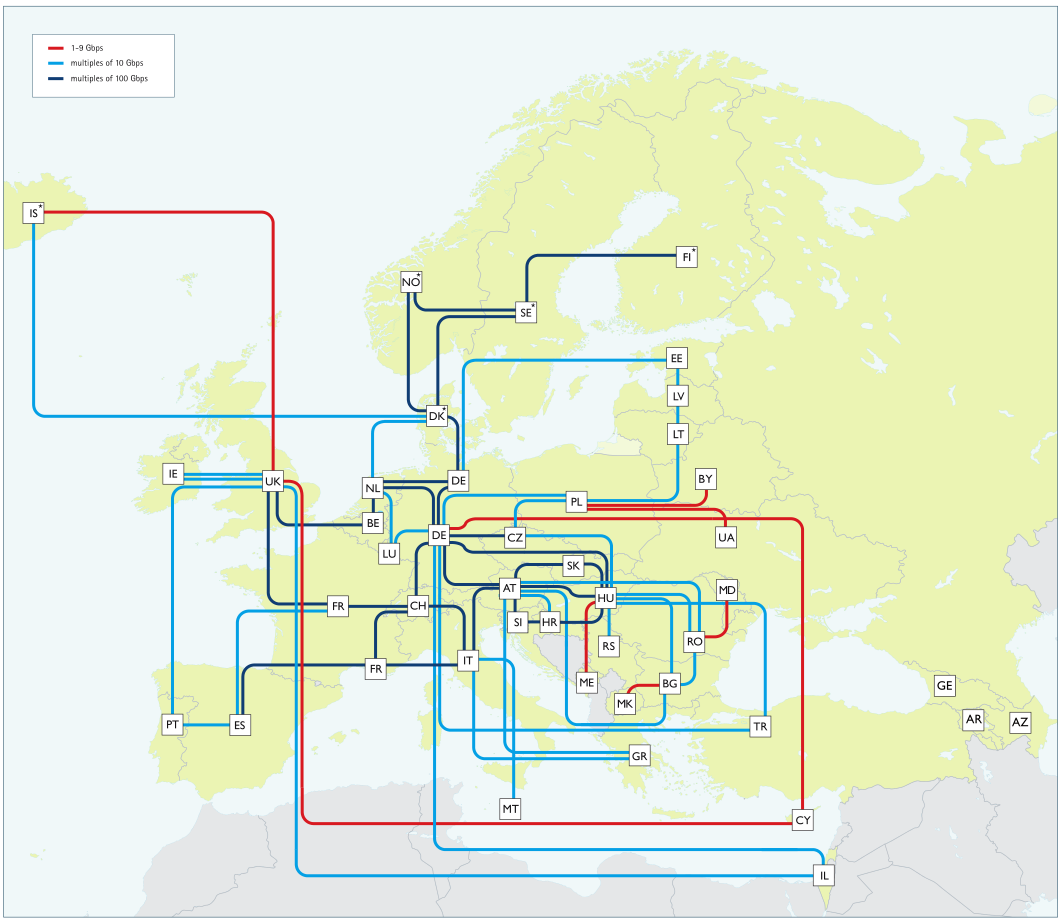}
\caption{G\'EANT topology (from \cite{GEANTTopology})}
\label{GEANT}
\end{figure}

To measure resource usage (i.e., computation time, number of flow entries, number of group tables and number of tags in use) and path lengths, it suffices to feed the application a topology instead of connecting it to a network. We evaluated the algorithm on the (high-level) network topology of the G\'EANT pan-European network (see Figure \ref{GEANT}), where we consider the 2 links between Ireland and the UK as 1 link.

We consider every node in the network to be a separate switch with one attached host 
and constructed the multicast tree with the host at Austria (AT) as source and all the other hosts as subscribers, by adding them to the multicast group one by one. This was repeated 5 times in a different order for all combinations of tree construction algorithm (SPT and DST) and fault tolerance requirement (single-link or 3-link fault tolerance). Austria was chosen as source, since it has a relatively large number of links (9) and is geographically central in the network. The computations were done on a Virtual Machine running Ubuntu 14.04.4 on an Intel i5-4670K CPU.

To measure the recovery times of both algorithms and compare them to the recovery time of the existing fast tree switching method we simulated a subset of the above G\'EANT network in Mininet. This subset consisted of switches AT, BE, CH, CY, both DE nodes, DK, ES, both FR nodes, IL, IS, IT, LU, NL, PT and UK. Hosts were only added to switches NL, UK and ES. To make sure ES was reachable by 4 disjoint paths, an additional link was added between ES and IT. All switches were simulated using Open vSwitch 2.4.0 \cite{OpenVSwitch}.

The host in NL would send packets at approximately $1/120$-second intervals, while the other two hosts would connect to their group and record all received packets. By cutting links and counting the missing packets at the receivers, we were able to get an approximation of the recovery times.

One of the main points in which our methods (and FF in general) improve upon methods that depend on the controller for failure recovery is that the recovery time is independent of the latency between the controller and the switches. As our controller application is executed on the same machine as the network itself is simulated, there is no such latency in our simulation, so to measure the effects of latency on recovery times we add an artificial delay of 10 ms between the controller and the switches using netem. This means that the round trip time is 20 ms, so this should be a minimum on the recovery time using the fast tree switching approach.

As we measure recovery time on a simulated network instead of on physical switches, the result is not completely accurate, but it provides a good indication as long as we take into account that:
\begin{itemize}
\item Flow entries will be installed more quickly than they would on physical switches \cite{huang2013high}. This has 2 major effects: the initial delay between subscribing and receiving the first packet will be lower than on a physical network and the recovery time of the fast tree switching approach will be lower than on a physical network.
\item The controller, the network simulation and the hosts are all running on the same machine, thus they take up each others processing time. To mitigate this, we used a subset of the G\'EANT topology instead of simulating all switches and hosts we use in the resource usage experiments. Unfortunately this does not completely solve the problem. One of the effects of this is that the source host sometimes sends a packet too late. To compensate for this it sends the next packet(s) earlier. The end result is that the average interval between packets is $1/120$ seconds.
\item The latency between the hosts is inaccurate. Links have no latency and infinite bandwidth, while switches use a software table to match flow entries instead of a hardware table.
\end{itemize}

In addition, we let links fail by turning off their interfaces directly. This results in a detection time of almost 0 ms. This does not matter when comparing different methods, but does mean that, depending on the failure detection method used, some time should be added to our values to get a more realistic recovery time. Assuming the use of BFD with a transmit interval of 1 ms and a multiplier of 3, around 4-5 ms should be added to our recovery times.

We only measured recovery time using the SPT algorithm, as using SPT or DST should not have a large effect on recovery time.

Recovery time was measured both for a single-link fault tolerance scenario, by setting $F$ to 1 and dropping 1 link, and for a three-link fault tolerance scenario, by setting $F$ to 3 and dropping 3 links. Our single-link scenario was as follows:

(1) Start transmitting from NL, (2) Connect UK to multicast group, (3) Wait 1 second, (4) Connect ES to multicast group, (5) Wait 10 seconds, (6) Drop link between NL and BE, (7) Wait 60 seconds, (8) Experiment is finished.

The three-link failure scenario was similar:

(1) Start transmitting from NL, (2) Connect UK to multicast group, (3) Wait 1 second, (4) Connect ES to multicast group, (5) Wait 10 seconds, (6) Drop links between NL and BE, IS and UK, and NL and DE2, (7) Wait 60 seconds, (8) Experiment is finished.

\subsection{Results}

\subsubsection{Computation Times}
\begin{table}[!t]
\renewcommand{\arraystretch}{1.3}
\caption{Primary path computation times}
\label{table_primarypathtimes}
\centering
\begin{tabular}{c|c|c|c}
\hline
\bfseries SPT (F=1) & \bfseries DST (F=1) & \bfseries SPT (F=3) & \bfseries DST (F=3)\\
\hline\hline
0.279 ms & 0.292 ms & 0.297 ms & 0.310 ms \\
\hline
\end{tabular}
\end{table}
\begin{figure*}[!t]
\centering
\subfloat{\includegraphics[width=0.5\linewidth]{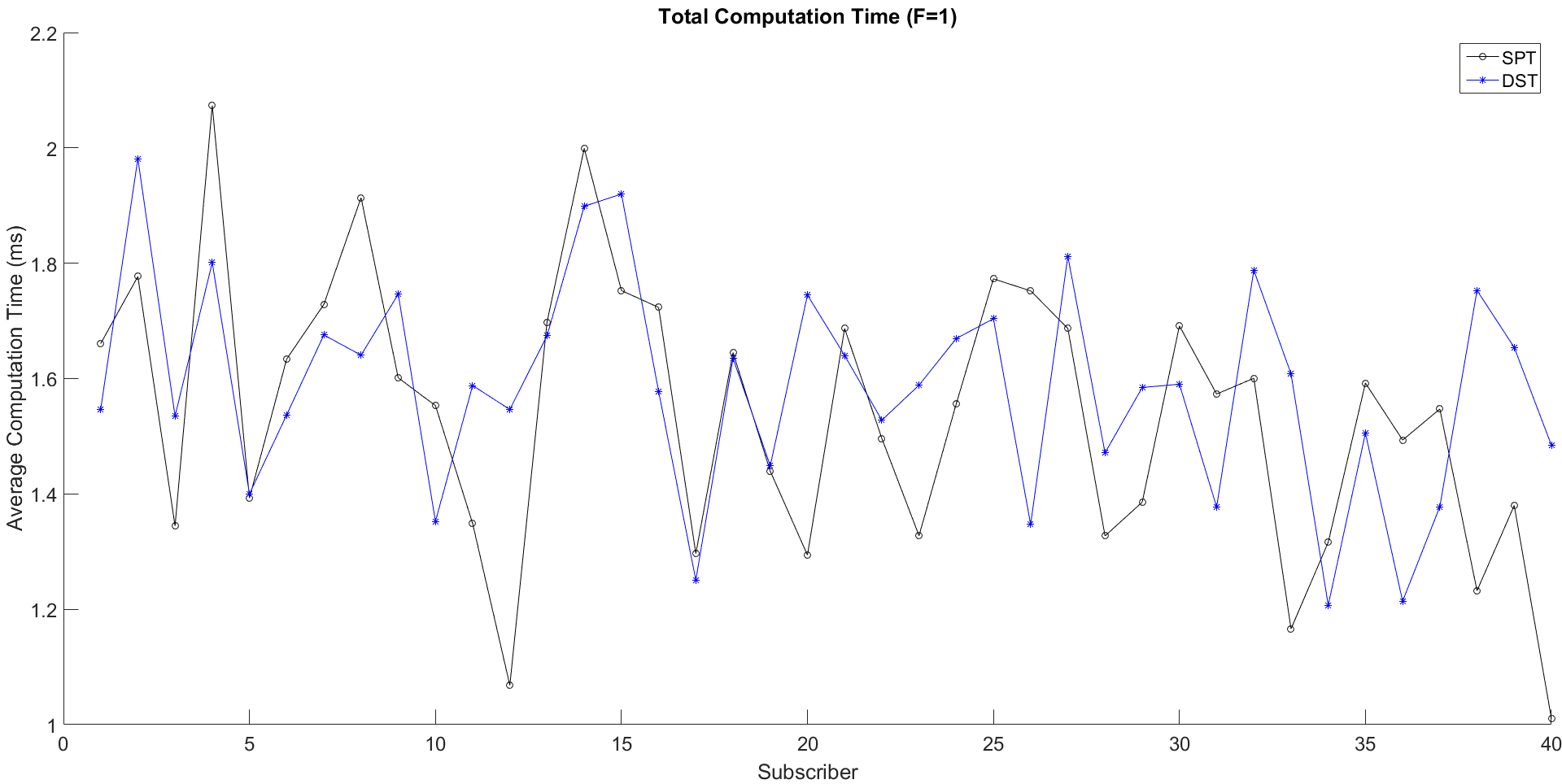}}
\hfil
\subfloat{\includegraphics[width=0.5\linewidth]{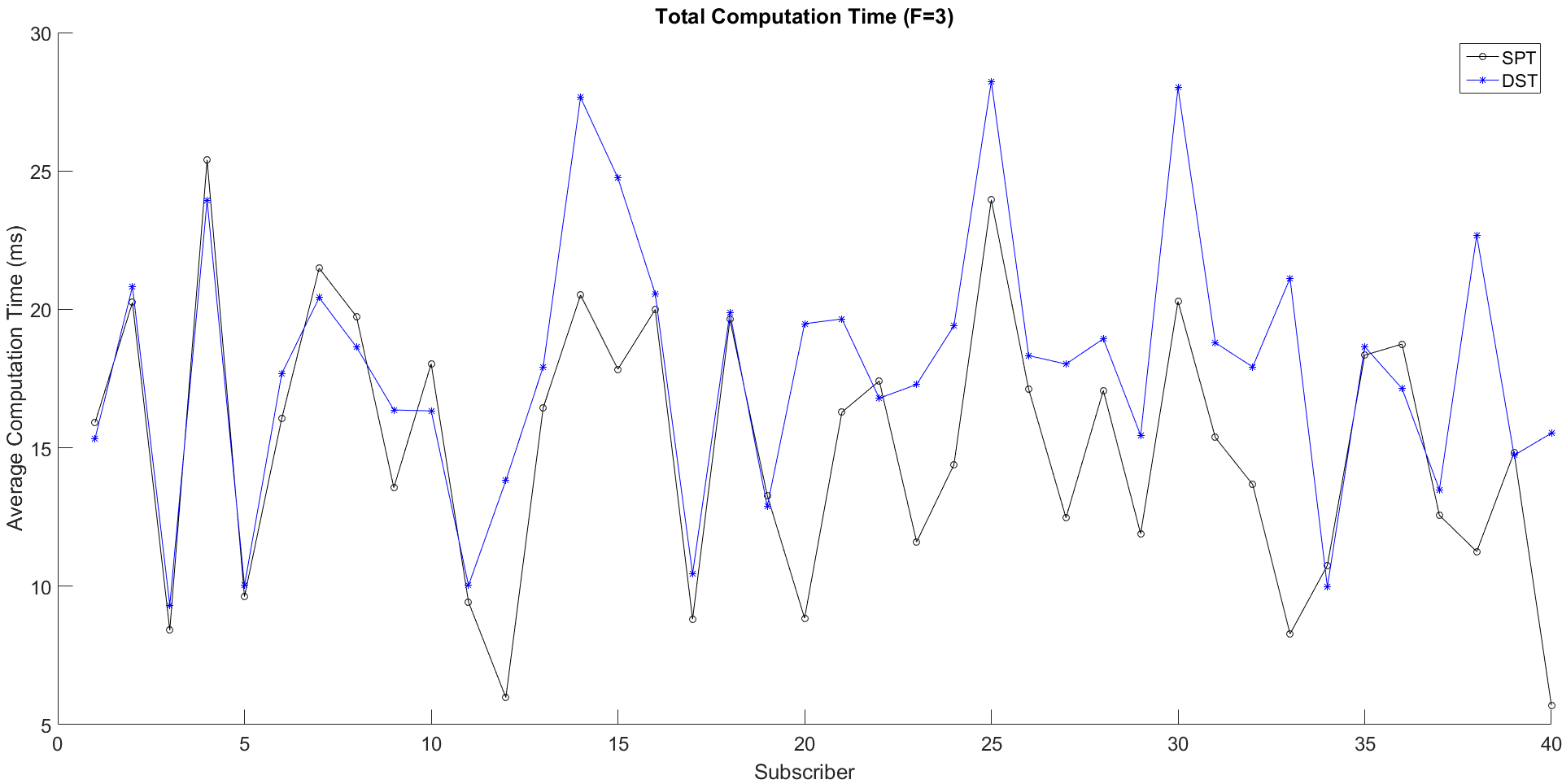}}
\caption{Average computation time of primary and backup paths}
\label{fig_TotalComputation}
\end{figure*}

The computation time of the algorithm is not only important because it takes up valuable processing time of the controller, but also because it influences the amount of time required till the first packet reaches a new subscriber. If a packet takes too long to reach the subscriber, the application might feel unresponsive to the end user.

As the primary path is computed and installed first, the amount of time until the first packet reaches the new subscriber is only influenced by the time it takes to compute this path, and not by the computation time of the backup paths. The average primary path computation times can be found in Table \ref{table_primarypathtimes}. These values are all well under a millisecond. We can conclude that, for the G\'EANT topology, the effect of the computation time is insignificant compared to the flow entry installation time and latency between the source and subscriber. This was to be expected, as calculating a single path only takes $O(|E| + |V| \log |V|)$ for both SPT and DST.

The total computation times can be found in Figure \ref{fig_TotalComputation}. In the single link case ($F=1$) all paths can almost always be computed within a few milliseconds. By increasing the fault tolerance requirement to 3 link failures, the computation time increases significantly.

\subsubsection{Flow Entries}
\begin{figure*}[!t]
\centering
\subfloat{\includegraphics[width=0.5\linewidth]{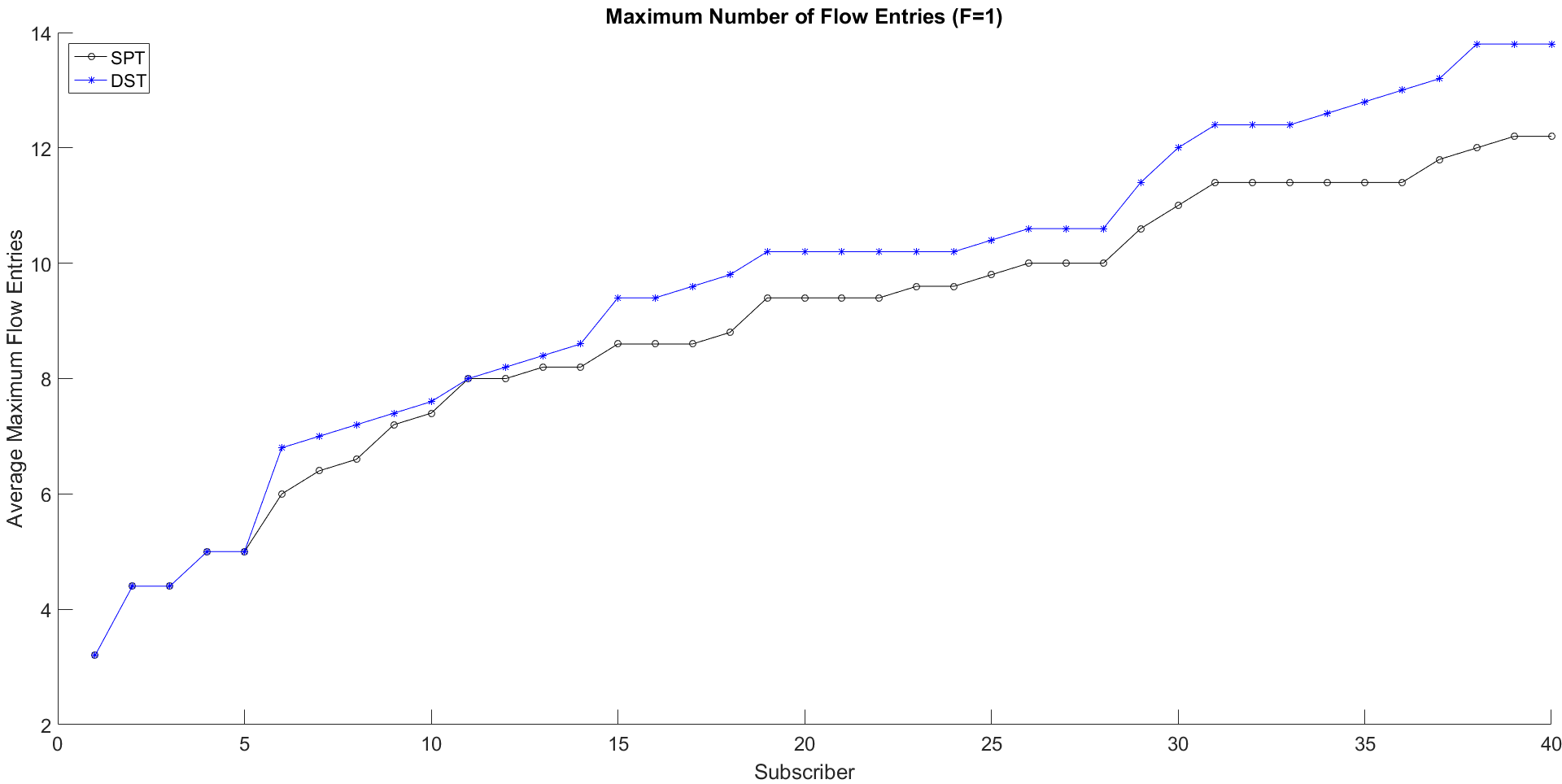}}
\hfil
\subfloat{\includegraphics[width=0.5\linewidth]{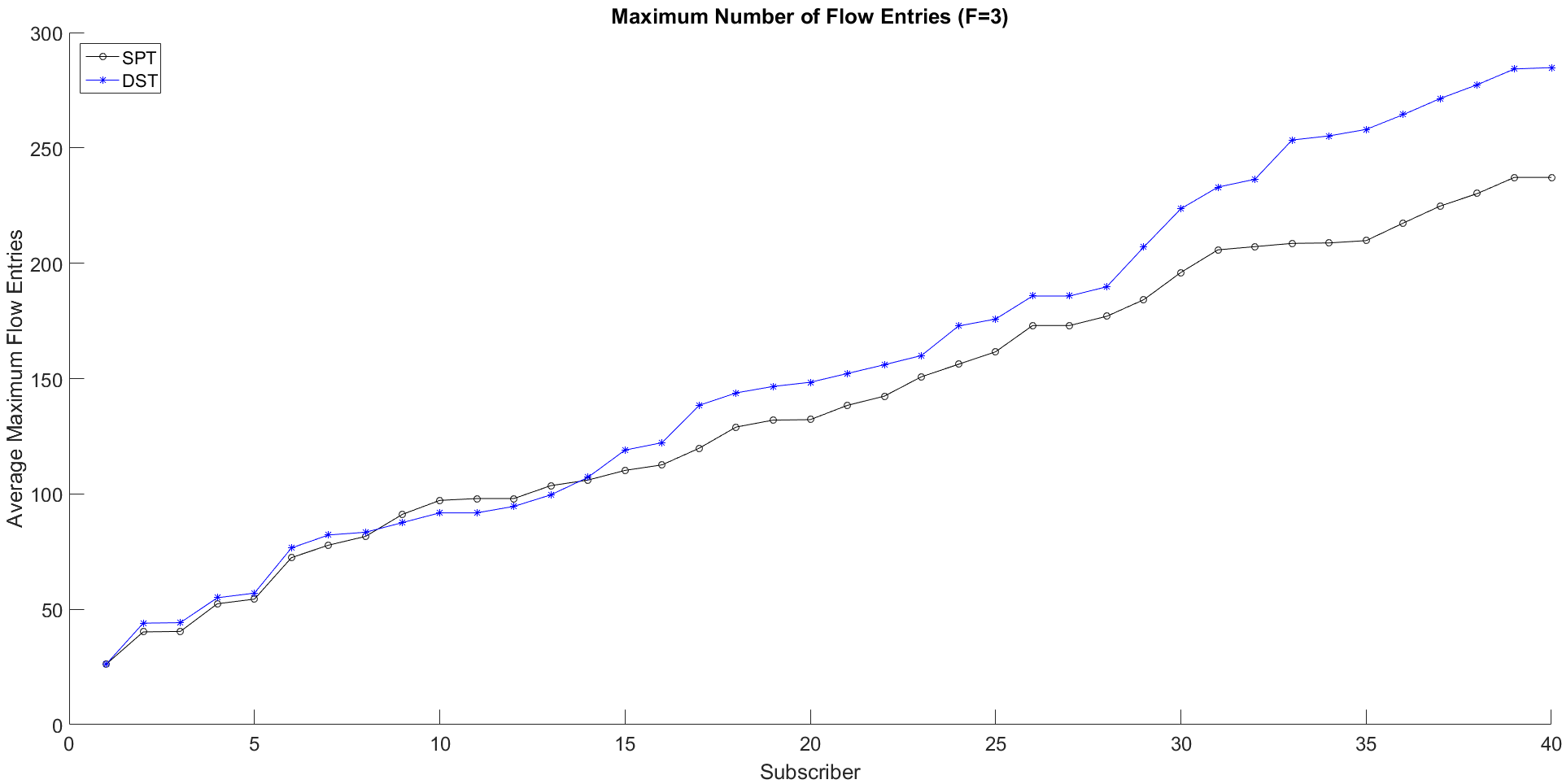}}
\caption{Average of the amount of flow entries in the switch with the most flow entries}
\label{fig_MaxFlows}
\end{figure*}
\begin{figure*}[!t]
\centering
\subfloat{\includegraphics[width=0.5\linewidth]{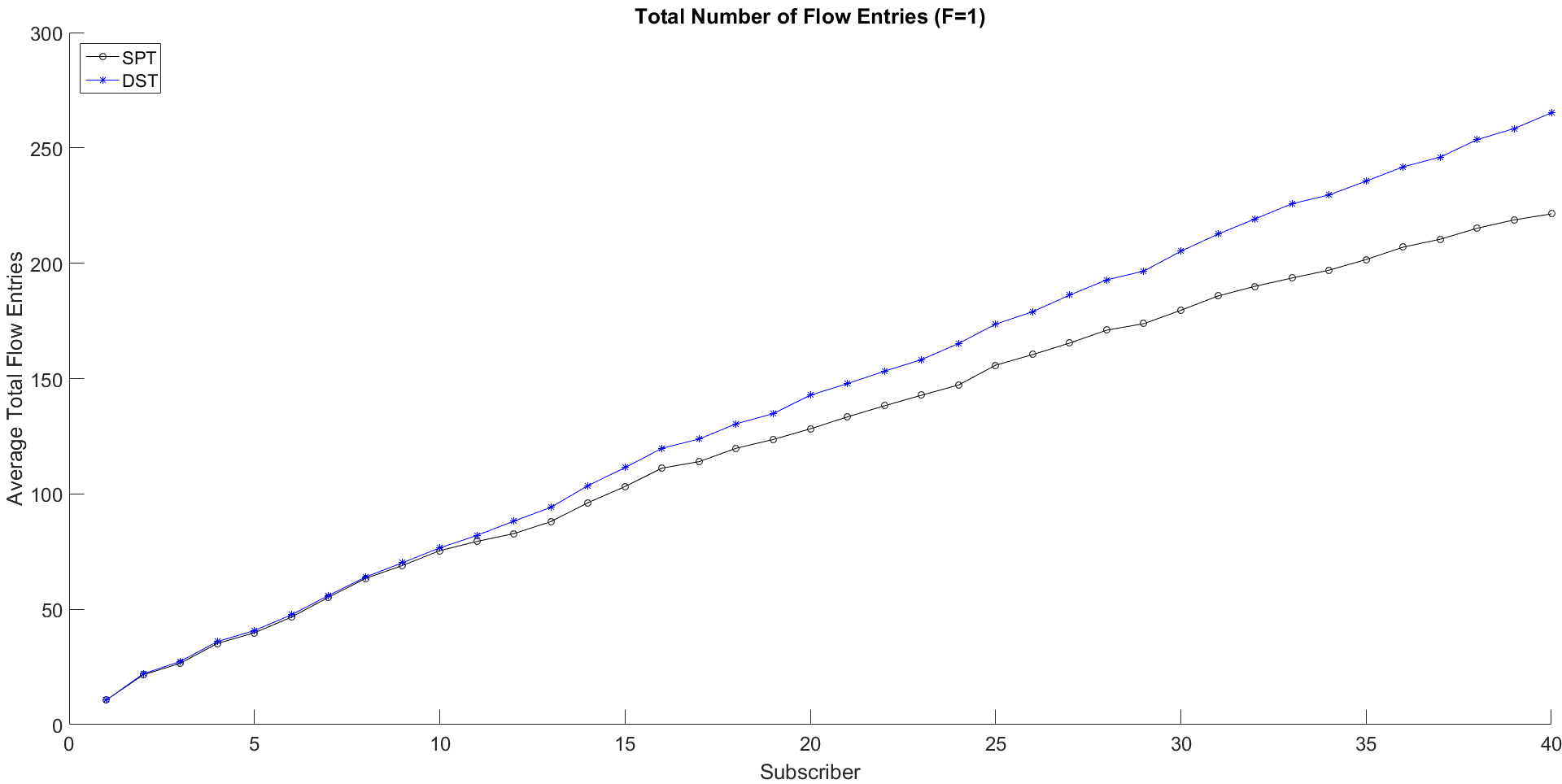}}
\hfil
\subfloat{\includegraphics[width=0.5\linewidth]{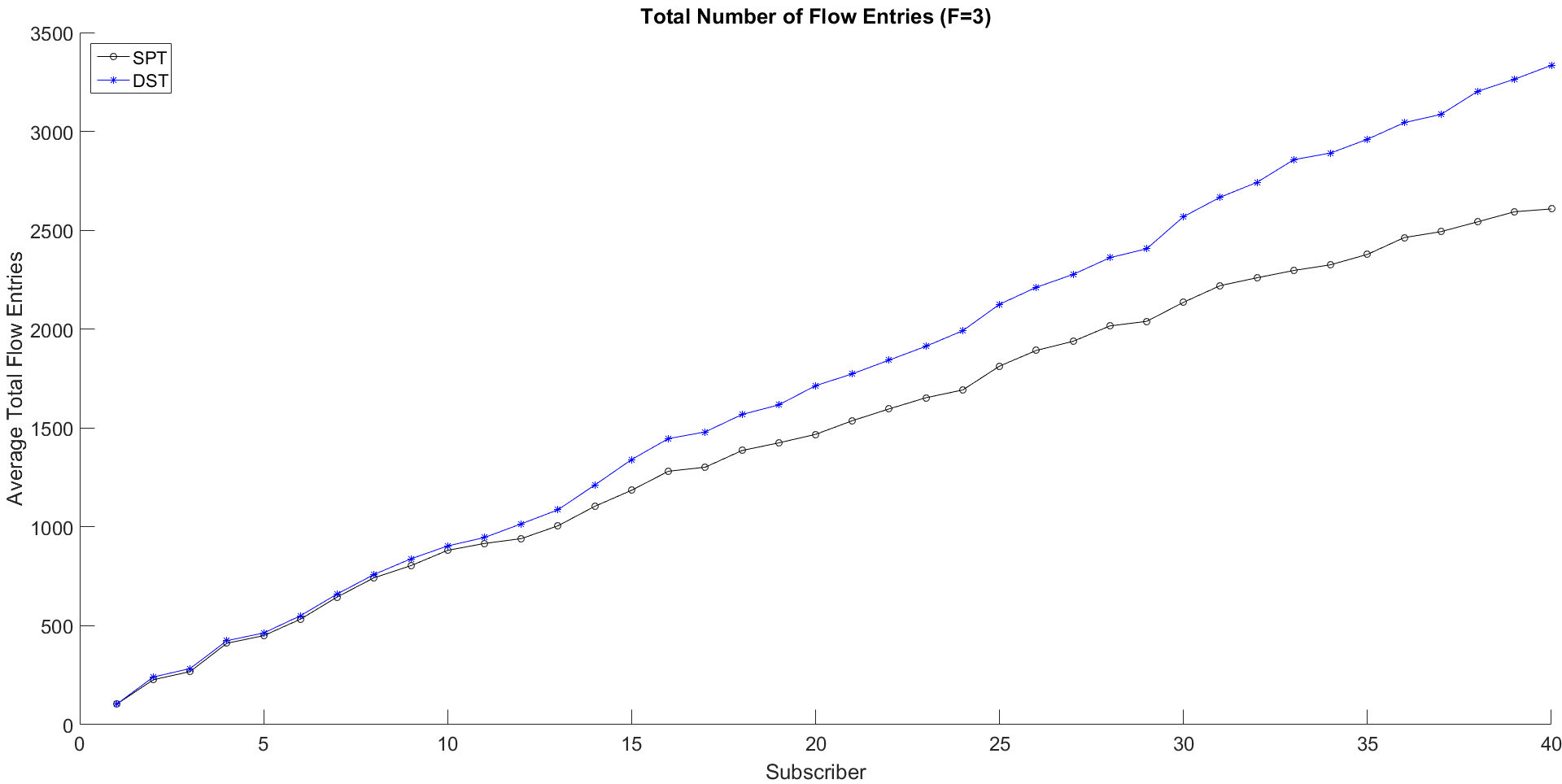}}
\caption{Average of the total amount of flow entries installed}
\label{fig_TotalFlows}
\end{figure*}

The amount of TCAM memory available in switches to install flow entries in is severely limited. Unfortunately, pre-installing backup paths for all possible link failures requires the installation of many flow entries. There are switches that also contain a software table, which can store many more flow entries, however, matching on this table is much slower than matching on the hardware (TCAM) flow table.

As can be seen in Figures \ref{fig_MaxFlows} and \ref{fig_TotalFlows}, in the single link fault tolerance case both the maximum amount of flow entries installed on a single switch as well as the total number of flow entries is low enough for the network to be capable of supporting multiple multicast groups. Unfortunately, when requiring more fault tolerance, the number of installed flow entries increases drastically.

It is interesting that the Dynamic Steiner Tree algorithm leads to a higher number of flow entries installed than the Shortest Path Tree algorithm. Although it tries to minimize the amount of links in a single tree, in this case the backup paths for the DST approximation turn out to be longer than the backup paths for the SPT.

\subsubsection{Group Tables}
\begin{figure*}[!t]
\centering
\subfloat{\includegraphics[width=0.5\linewidth]{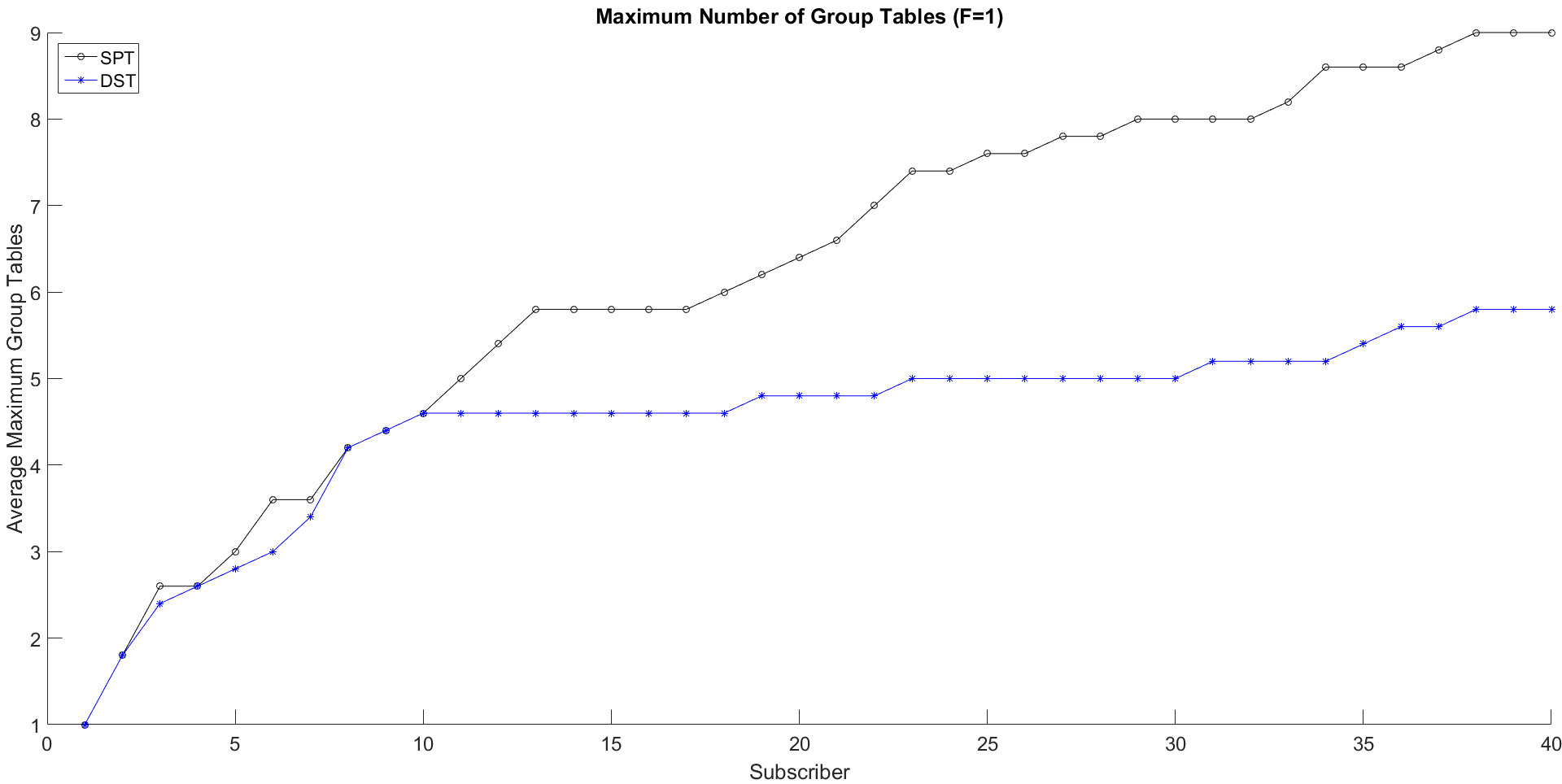}}
\hfil
\subfloat{\includegraphics[width=0.5\linewidth]{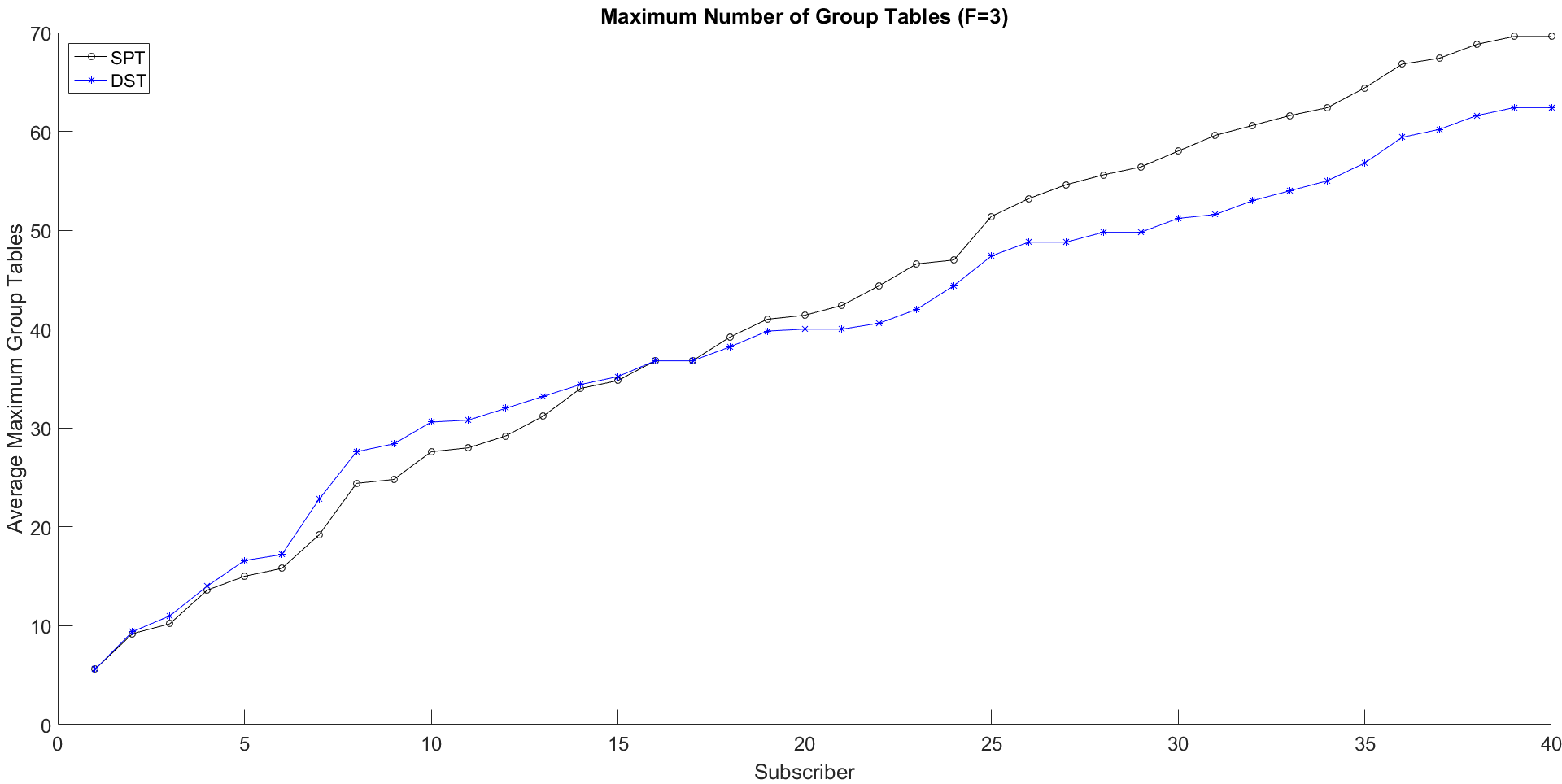}}
\caption{Average of the amount of group tables in the switch with the most group tables}
\label{fig_MaxGroups}
\end{figure*}
\begin{figure*}[!t]
\centering
\subfloat{\includegraphics[width=0.5\linewidth]{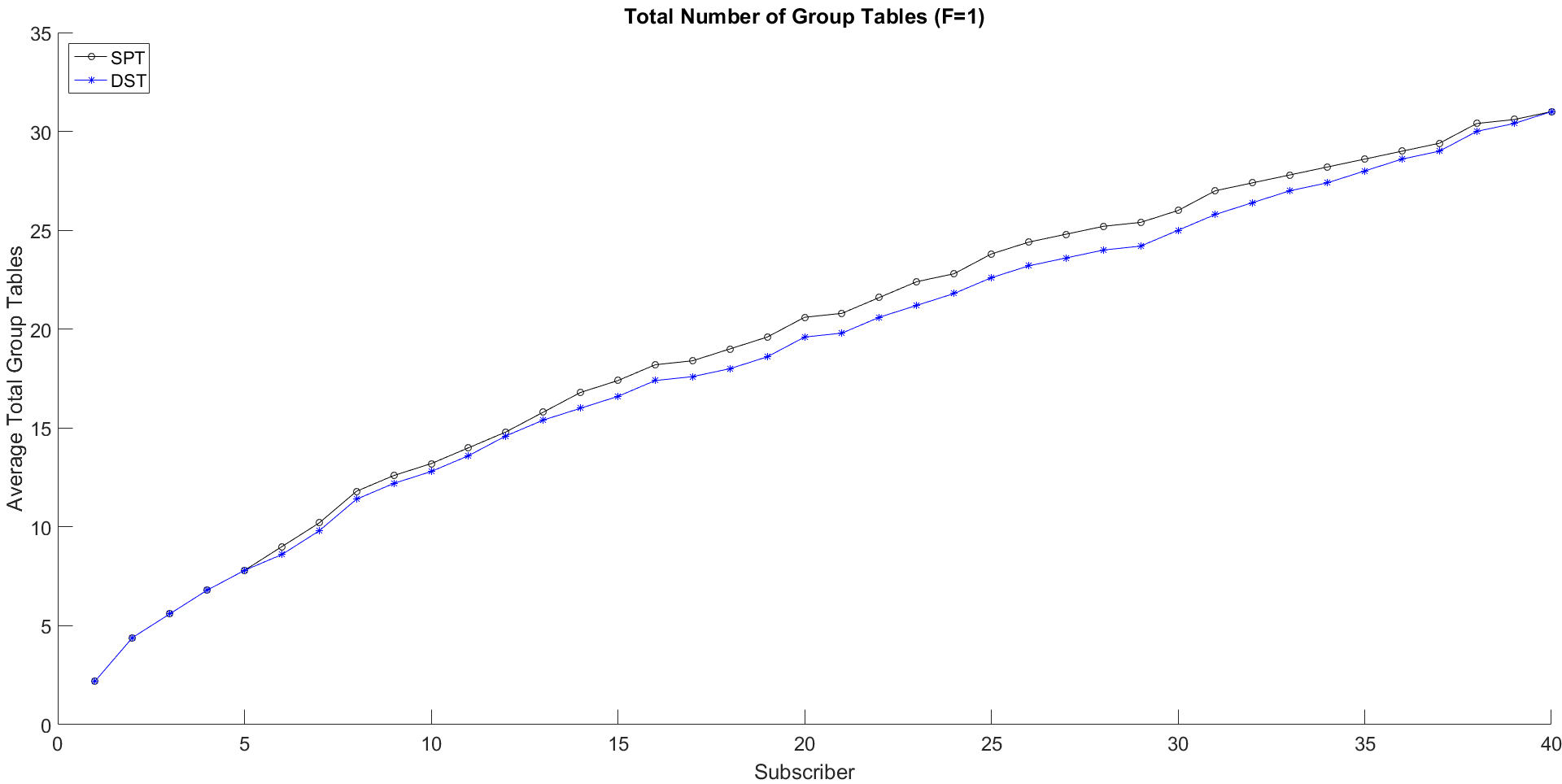}}
\hfil
\subfloat{\includegraphics[width=0.5\linewidth]{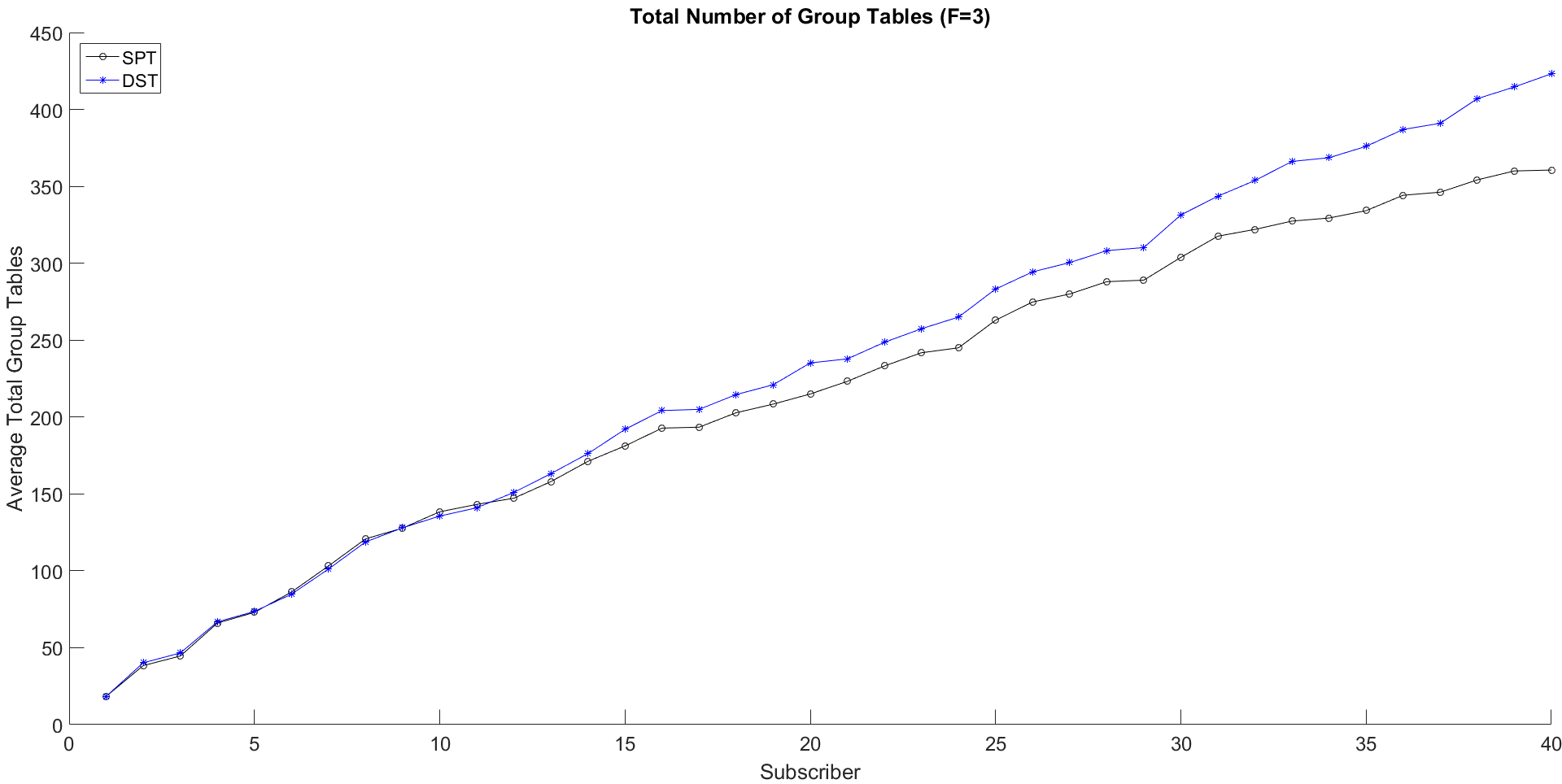}}
\caption{Average of the total amount of group tables in use}
\label{fig_TotalGroups}
\end{figure*}
While the amount of flow entries that can be installed in an OpenFlow switch is quite low, the amount of group tables that can be used is even lower. As such, it is of vital importance to limit the amount of group tables taken up by the multicast groups.

The average amount of group tables in use is plotted in Figures \ref{fig_MaxGroups} and \ref{fig_TotalGroups}.

The number of group tables used in the single-link fault tolerance case is very small, but the amount of group tables used in the $F=3$ case is too high. Consider for example the HP 2920 Switches, these switches only support 32 group tables per OpenFlow instance \cite{HPGuide}. On these switches, one could only add around 12 subscribers to the multicast group (and not support any more multicast groups).

Note that in this case it might be advantageous to use the DST approximation algorithm, as this results in a slight decrease in the maximum amount of group tables in use.

\subsubsection{Tags}
\begin{figure*}[!t]
\centering
\subfloat{\includegraphics[width=0.5\linewidth]{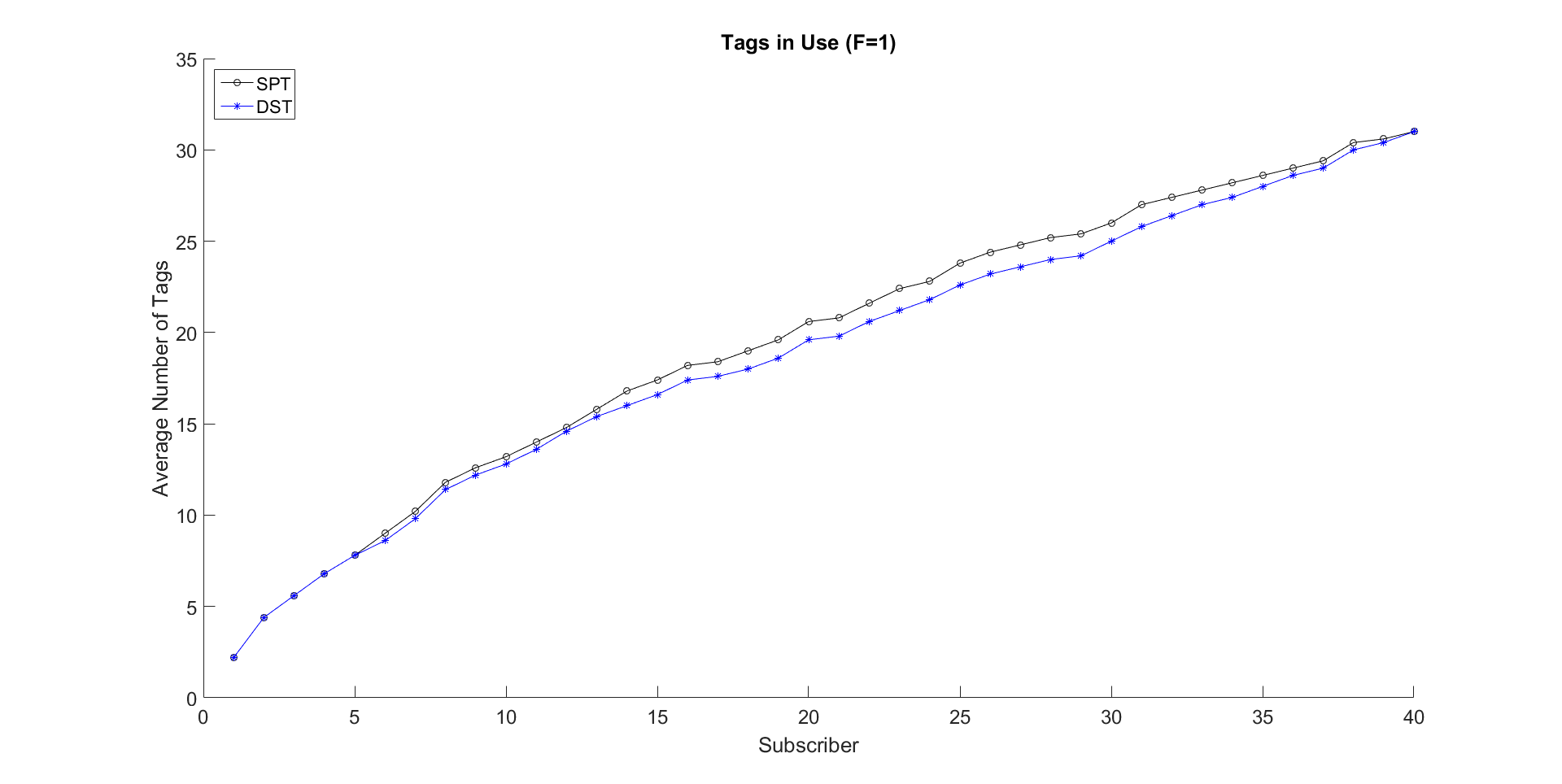}}
\hfil
\subfloat{\includegraphics[width=0.5\linewidth]{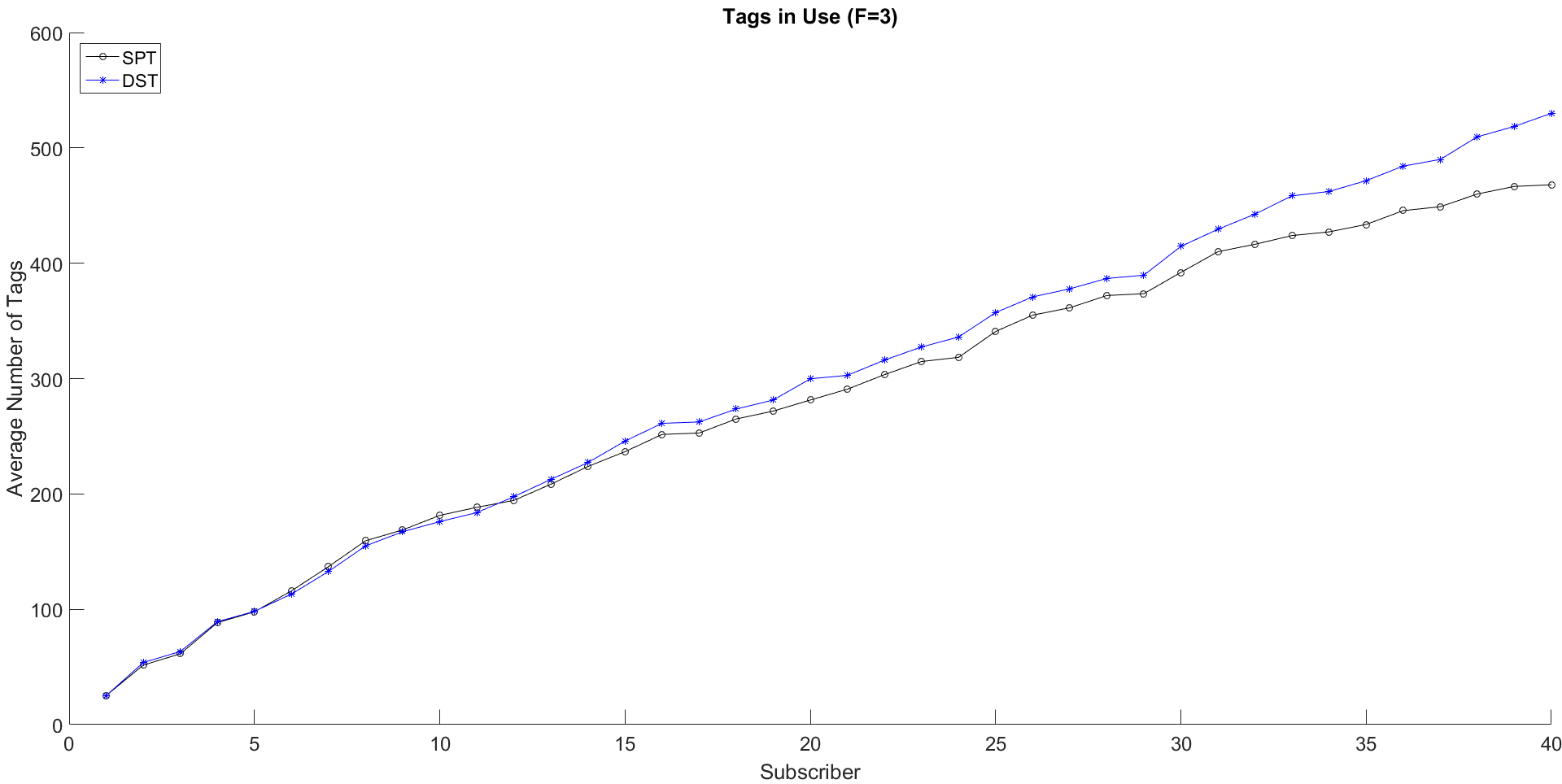}}
\caption{Average of the amount of tags in use}
\label{fig_Tags}
\end{figure*}

There are only 4,094 VLAN tags available on a given network. However, multiple groups can share their tags, as tags only need to be unique for a single source+group combination. In addition, if more tags are required, MPLS labels can be used instead of VLAN tags.

As can be seen in Figure \ref{fig_Tags} there are more than enough VLAN tags available to support 3 link fault tolerance.

\subsubsection{Path Lengths}
\begin{table}[!t]
\renewcommand{\arraystretch}{1.3}
\caption{Average hopcounts from source to subscriber for different number of link failures}
\label{Hopcounts}
\centering
\begin{tabular}{c||c|c|c|c}
\hline
 & \bfseries 0 links & \bfseries 1 link & \bfseries 2 links & \bfseries 3 links\\
\hline\hline
\bfseries SPT & 2.3750 & 3.4479 & 4.6546 & 5.9739 \\
\bfseries DST & 3.2300 & 4.0620 & 5.3421 & 6.6414 \\
\hline
\end{tabular}
\end{table}

In Table \ref{Hopcounts}, we compare the average path length between source and subscriber for different amounts of failing links. As can be expected, constructing shortest path trees results in the lowest average hopcount between source and subscribers.

\subsubsection{Recovery Time}
\begin{table*}[!t]
\renewcommand{\arraystretch}{1.3}
\caption{Results from link failure experiments (10 ms latency between controller and switches)}
\label{table_failure}
\centering
\begin{tabular}{c||c|c|c|c}
& \bfseries Packets lost (UK) & \bfseries Packets lost (ES) & \bfseries Initial delay (UK) & \bfseries Initial delay (ES)\\ \hline \hline
\bfseries FF (1 link failure) & 0 & 0 & 28.457 ms & 37.073 ms\\
\bfseries Fast tree switching (1 link failure)& 3 & 3 & 26.693 ms & 19.317 ms\\ \hline
\bfseries FF (3 link failure) & 0 & 0 & 56.820 ms & 74.105 ms\\
\bfseries Fast tree switching (3 link failure)& 65 & 65 & 34.252 ms & 40.433 ms\\
\end{tabular}
\end{table*}
\begin{table*}[!t]
\renewcommand{\arraystretch}{1.3}
\caption{Results from link failure experiments (no latency between controller and switches)}
\label{table_failure_nolatency}
\centering
\begin{tabular}{c||c|c|c|c}
& \bfseries Packets lost (UK) & \bfseries Packets lost (ES) & \bfseries Initial delay (UK) & \bfseries Initial delay (ES)\\ \hline \hline
\bfseries Fast tree switching (1 link failure)& 0 & 0 & 5.472 ms & 8.405 ms\\ \hline
\bfseries Fast tree switching (3 link failure)& 0 & 0 & 23.195 ms & 14.287 ms\\
\end{tabular}
\end{table*}

Table \ref{table_failure} shows the results of our recovery time experiments. Using the FF algorithm no packets are lost. This is exactly what we would expect, as the recovery time should be about equal to the detection time, which is zero in our experiments. The fast tree switching algorithm does lose multiple packets, and especially loses many packets when multiple links fail at the same time.

We would expect the delay between subscribing and receiving the first packet (initial delay) to not increase when protecting the network against more failures, as the primary path is computed and installed before all backup paths. Yet, Table \ref{table_failure} shows an increase. This is likely due to the longer processing time of the controller (taking up time switches would otherwise use to route traffic). 

Based on the packet loss and the packet rate (120 per second) we estimate the recovery time of Fast Failover to be approximately $0 - 8.333$ ms in both fault tolerance cases and the recovery time of fast tree switching to be around $16.666 - 33.333$ ms for the single-link case and around $533.333 - 550$ ms for the three-link case. 

Note that in our experiment only a single multicast group needed to be recovered. Based on the results in \cite{kotani2012design,DaisukeKotani2016} we expect the recovery time for the fast tree switching method to increase significantly with increasing number of groups. Additionally, installing flow entries takes more time on physical switches than it takes on software switches, which further increases the recovery time on a non-simulated network. In contrast, the recovery time of FF will not be affected by both the number of multicast groups or the flow entry installation time.

To demonstrate that the packet loss of the fast tree switching method is mostly caused by the latency between the controller and the switches, we also performed two additional experiments where no latency was added. The results of these experiments can be found in Table \ref{table_failure_nolatency}. When no latency was added between the controllers and switches, the fast tree switching algorithm achieved a packet loss of $0$.

\subsubsection{Complete Graph}
\begin{table}[!t]
\renewcommand{\arraystretch}{1.3}
\caption{Primary path computation times on the complete graph}
\label{table_primarypathtimesC}
\centering
\begin{tabular}{c|c|c|c}
\hline
\bfseries SPT (F=1) & \bfseries DST (F=1) & \bfseries SPT (F=3) & \bfseries DST (F=3)\\
\hline\hline
1.771 ms & 1.744 ms & 1.743 ms & 1.774 ms \\
\hline
\end{tabular}
\end{table}
\begin{figure*}[!t]
\centering
\subfloat{\includegraphics[width=0.5\linewidth]{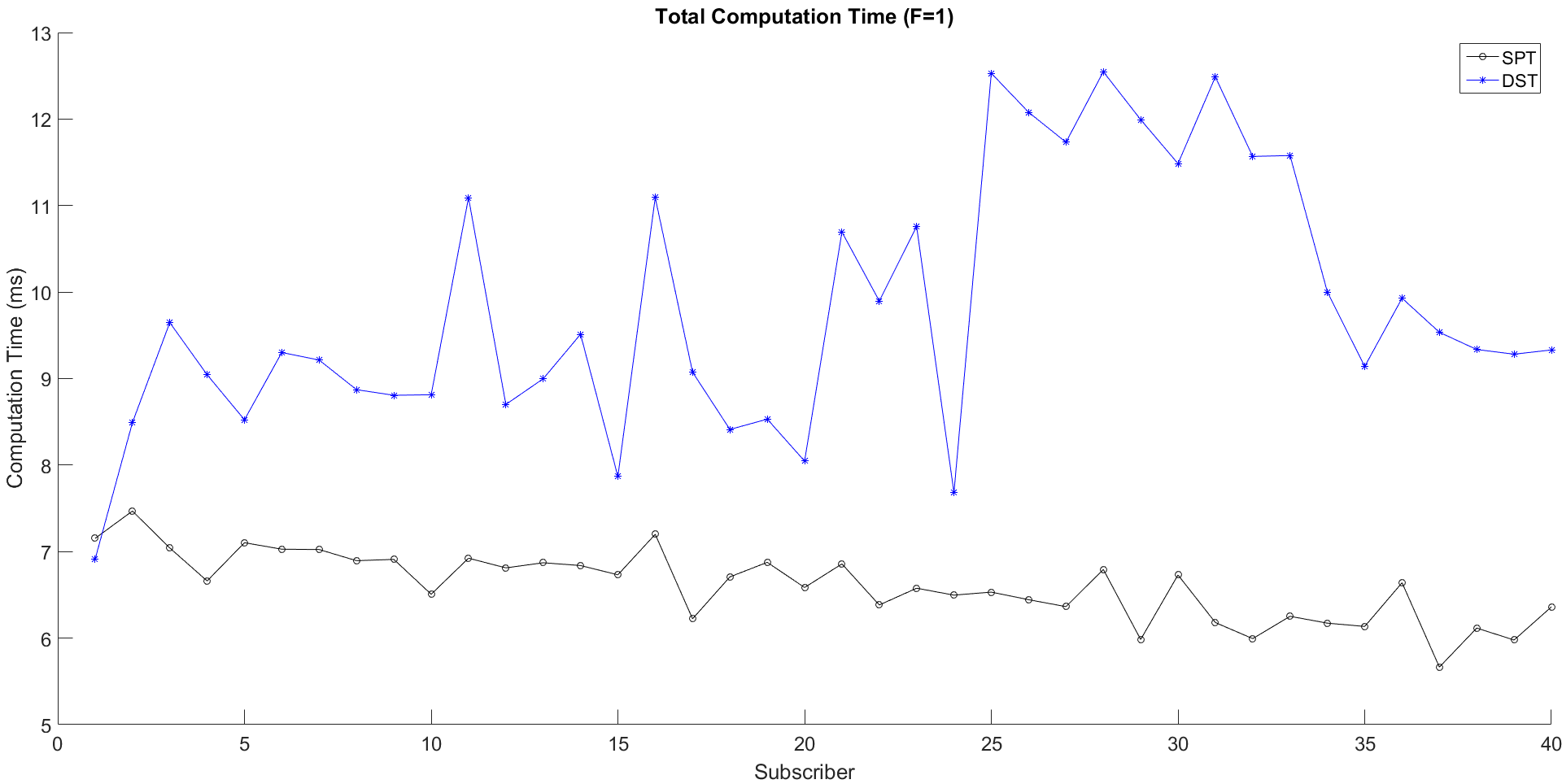}}
\hfil
\subfloat{\includegraphics[width=0.5\linewidth]{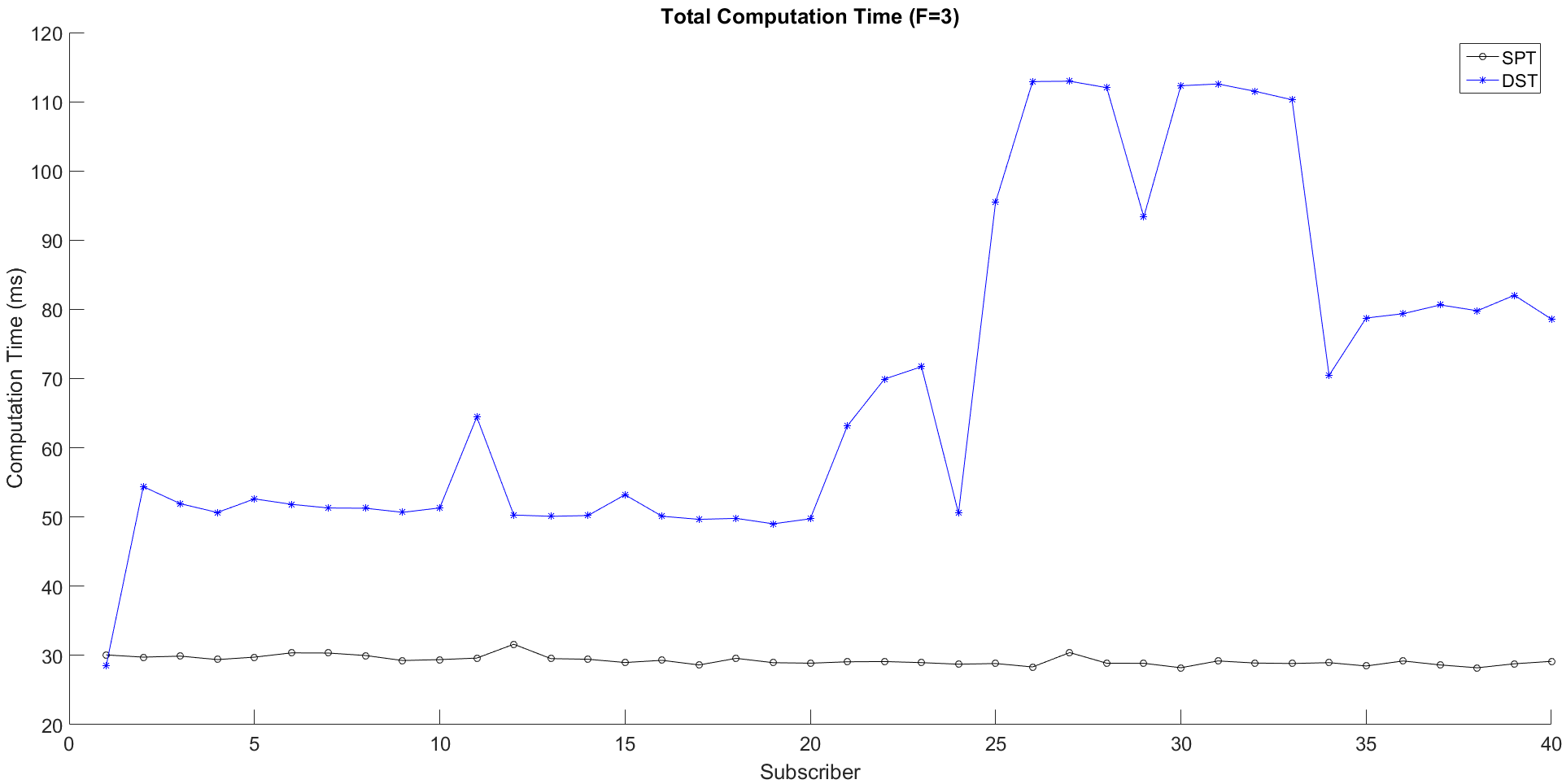}}
\caption{Average computation time of primary and backup paths on the complete graph}
\label{fig_TotalComputationC}
\end{figure*}

The complete graph is an interesting topology for measuring resource usage, as a maximum amount of disjoint backup paths are available and because path lengths will be as short as possible. 

As can be seen in Table \ref{table_primarypathtimesC} and Figure \ref{fig_TotalComputationC}, the computation time increases dramatically on the complete graph compared to the G\'EANT topology.

\begin{figure*}[!t]
\centering
\subfloat{\includegraphics[width=0.5\linewidth]{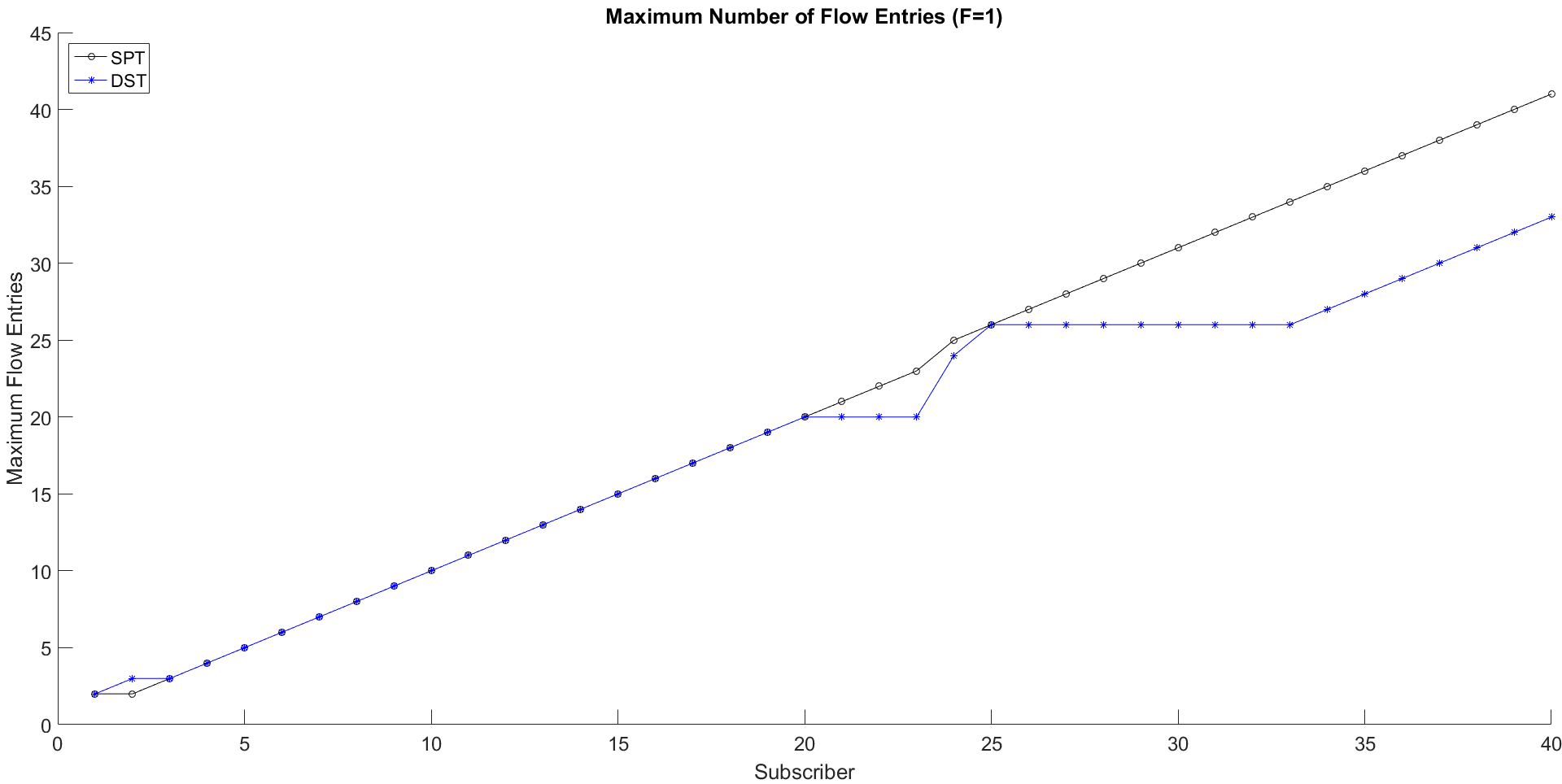}}
\hfil
\subfloat{\includegraphics[width=0.5\linewidth]{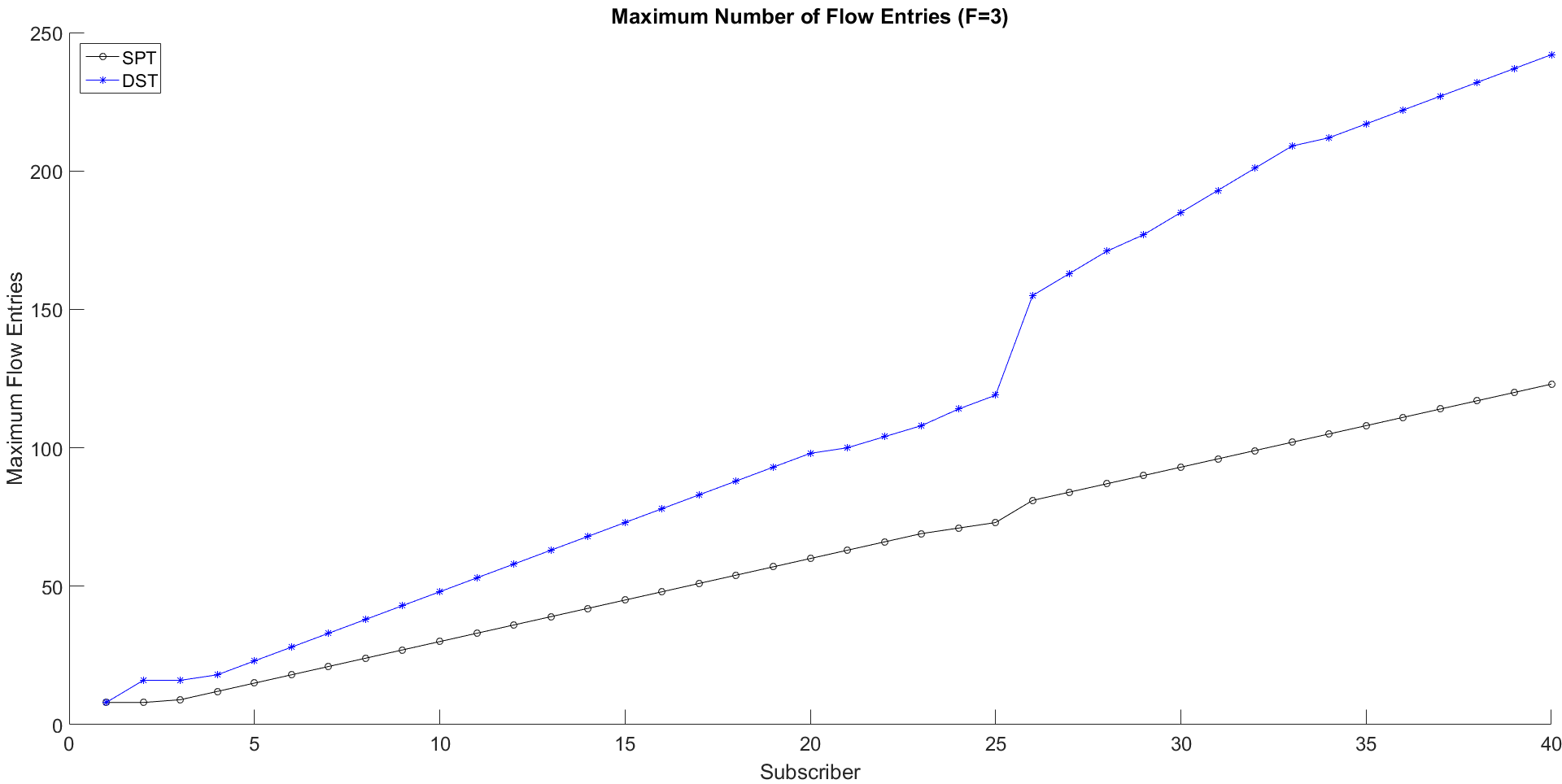}}
\caption{Average amount of flow entries in the switch with most flow entries on the complete graph}
\label{fig_MaxFlowsC}
\end{figure*}
\begin{figure*}[!t]
\centering
\subfloat{\includegraphics[width=0.5\linewidth]{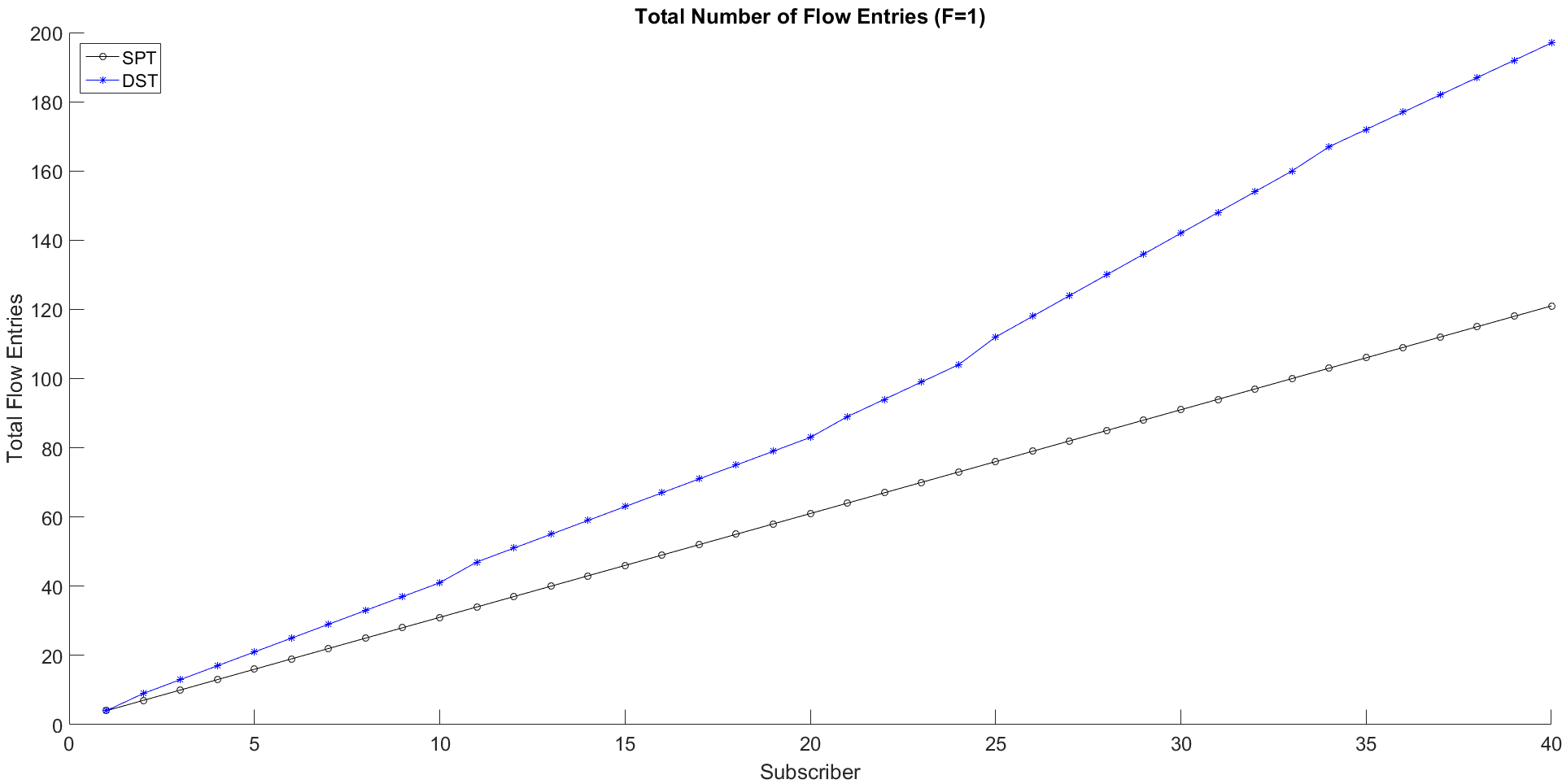}}
\hfil
\subfloat{\includegraphics[width=0.5\linewidth]{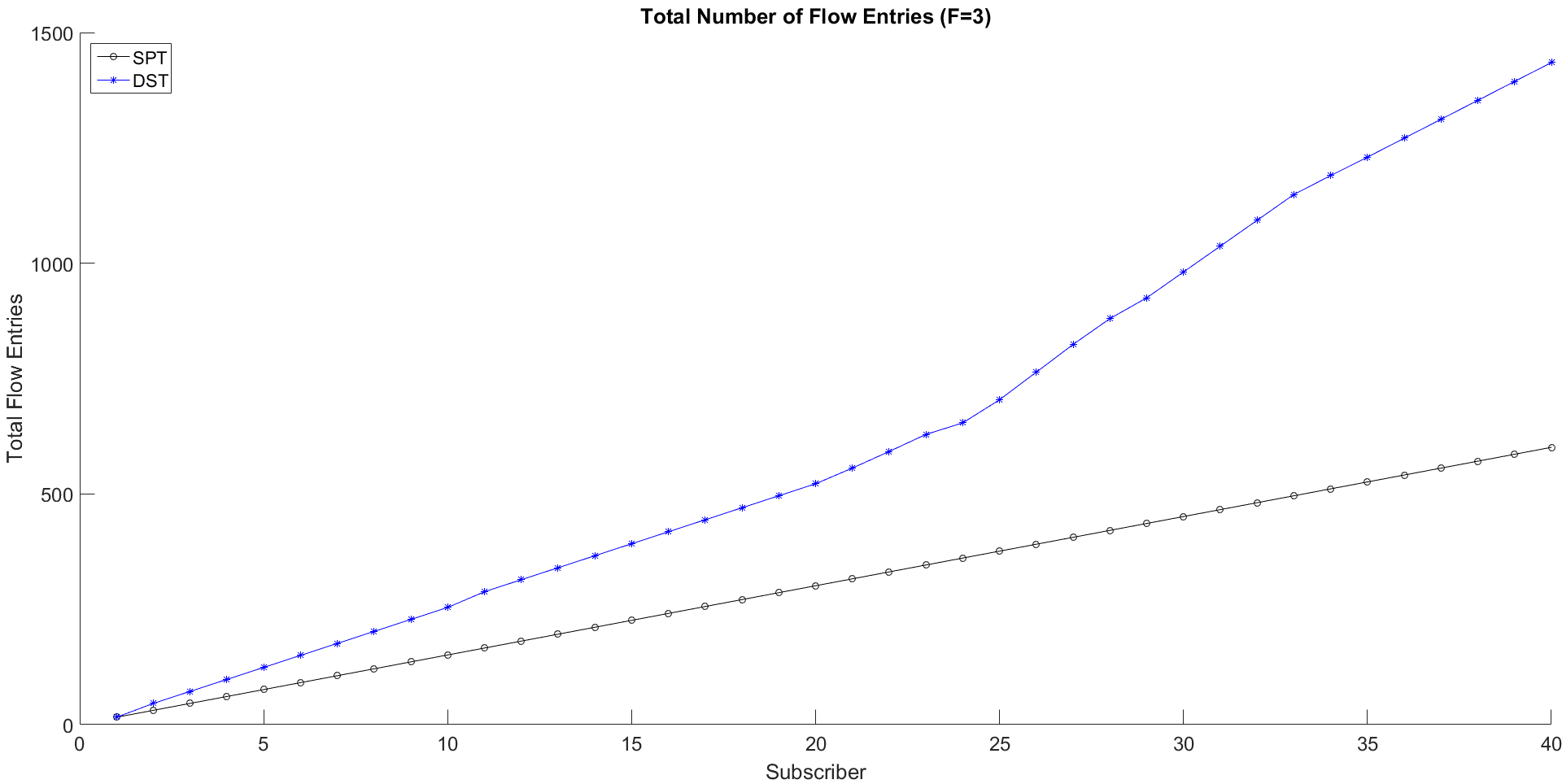}}
\caption{Average total amount of flow entries installed on the complete graph}
\label{fig_TotalFlowsC}
\end{figure*}

As could be expected, the total amount of flow entries is lower on the complete graph (Figure \ref{fig_TotalFlowsC}), but they are more concentrated in specific switches (Figure \ref{fig_MaxFlowsC}). Note that DST is not very efficient here, as it does not prioritize lower path lengths, even when comparing two possible paths that would both result in adding the same amount of links to the tree.

\begin{figure*}[!t]
\centering\includegraphics[width=0.5\linewidth]{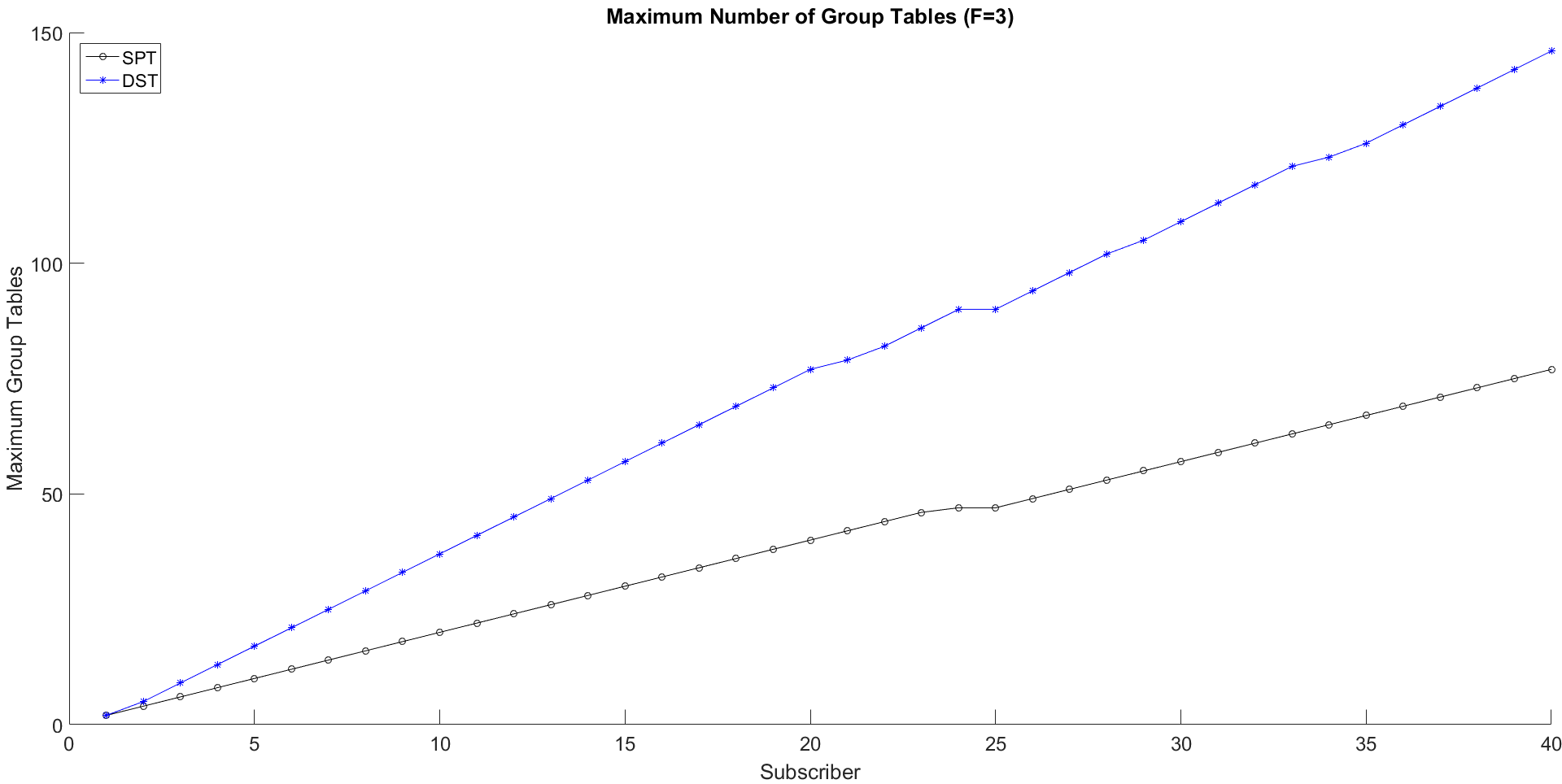}
\caption{Average amount of group tables in the switch with most group tables in the complete graph}
\label{fig_MaxGroupsC}
\end{figure*}
\begin{figure*}[!t]
\centering\includegraphics[width=0.5\linewidth]{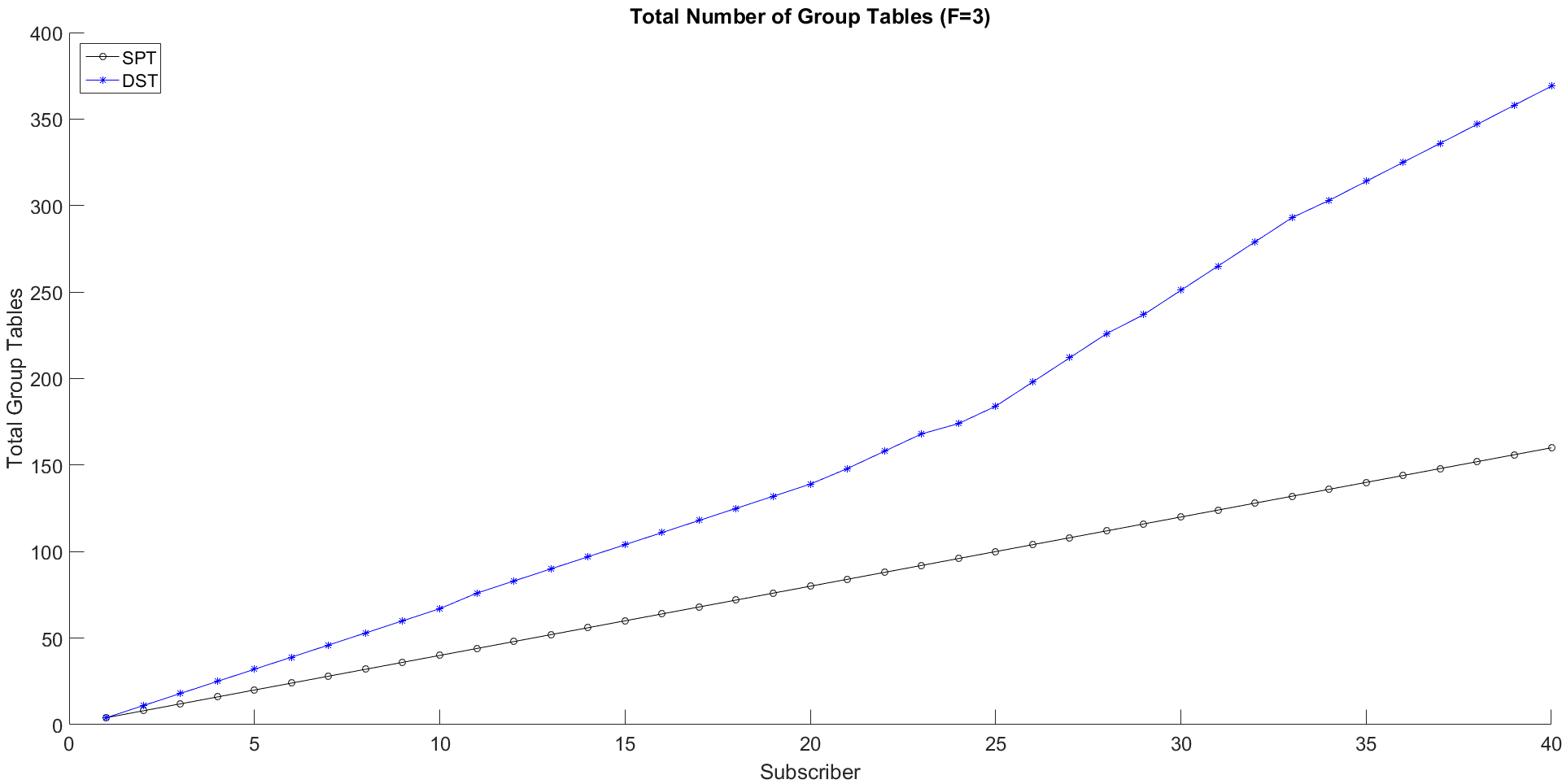}
\caption{Average total amount of group tables in use in the complete graph}
\label{fig_TotalGroupsC}
\end{figure*}

In the single-link fault tolerance case, one group table needs to be installed for every single subscriber, as they all require one additional link to be added to the tree. In the SPT case these group tables are all concentrated in the same switch, while in the DST case only 18 group tables at most are installed in a single switch.

The number of group tables installed in the three-link fault tolerance case can be seen in Figures \ref{fig_MaxGroupsC} and \ref{fig_TotalGroupsC}.

\begin{figure*}[!t]
\centering
\subfloat{\includegraphics[width=0.5\linewidth]{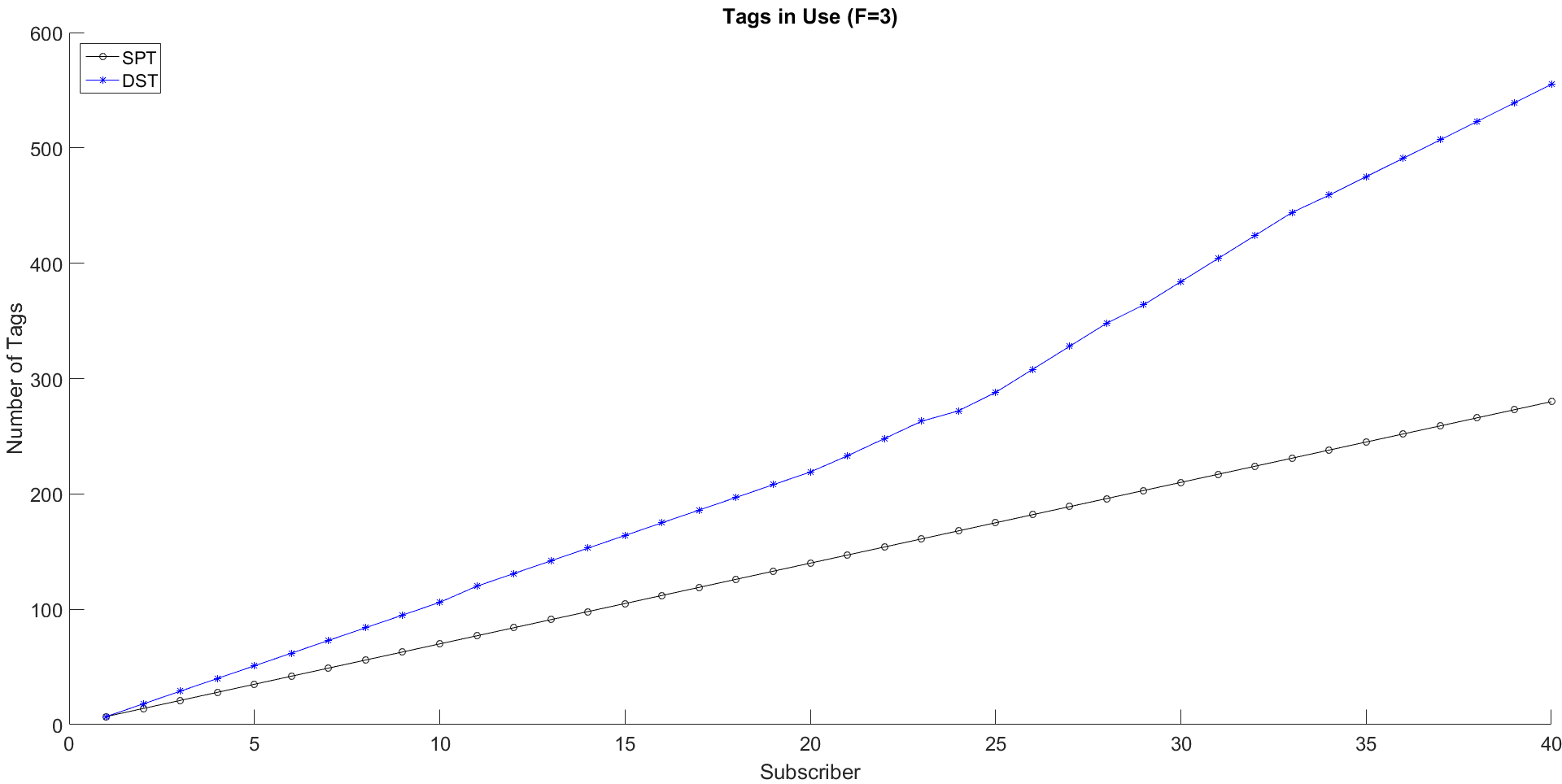}}
\caption{Average amount of tags in use in the complete graph}
\label{fig_TagsC}
\end{figure*}

The tag usage in the single-link fault tolerance case is exactly the same for all methods: 1 tag per subscriber. In Figure \ref{fig_TagsC}, we show the number of tags required in the $F=3$ case. It is clear that the number of tags is not an issue in the complete graph.

\begin{table}[!t]
\renewcommand{\arraystretch}{1.3}
\caption{Average hopcounts from source to subscriber for different number of link failures in the complete graph}
\label{HopcountsC}
\centering
\begin{tabular}{c||c|c|c|c}
\hline
 & \bfseries 0 links & \bfseries 1 link & \bfseries 2 links & \bfseries 3 links\\
\hline\hline
\bfseries SPT & 1.0000 & 2.0000 & 2.5000 & 3.0000\\
\bfseries DST & 2.7250 & 2.9229 & 3.4850 & 4.0130 \\
\hline
\end{tabular}
\end{table}

The average length of the paths between source and subscriber can be found in Table \ref{HopcountsC}.

\section{Conclusion}
\label{Conclusions}
We have proposed and evaluated an algorithm to provide fast failure recovery for dynamic multicast groups using Fast Failover tables. This algorithm can be used in combination with a variety of different tree construction algorithms to enable fault tolerance against $F$ link failures, where $F$ can be adjusted based on the needs of the streaming service. In contrast to other recovery methods, using Fast Failover groups uses up much more resources, both in terms of computation power in the controller, as in memory on the switches. However, FF also leads to a significant decrease in recovery time, especially when the round trip time between the controller and switches is larger than the failure detection time.

\bibliographystyle{IEEEtran}
\bibliography{IEEEabrv,bibfile}

\begin{thebibliography}{10}
\providecommand{\url}[1]{#1}
\csname url@samestyle\endcsname
\providecommand{\newblock}{\relax}
\providecommand{\bibinfo}[2]{#2}
\providecommand{\BIBentrySTDinterwordspacing}{\spaceskip=0pt\relax}
\providecommand{\BIBentryALTinterwordstretchfactor}{4}
\providecommand{\BIBentryALTinterwordspacing}{\spaceskip=\fontdimen2\font plus
\BIBentryALTinterwordstretchfactor\fontdimen3\font minus
  \fontdimen4\font\relax}
\providecommand{\BIBforeignlanguage}[2]{{%
\expandafter\ifx\csname l@#1\endcsname\relax
\typeout{** WARNING: IEEEtran.bst: No hyphenation pattern has been}%
\typeout{** loaded for the language `#1'. Using the pattern for}%
\typeout{** the default language instead.}%
\else
\language=\csname l@#1\endcsname
\fi
#2}}
\providecommand{\BIBdecl}{\relax}
\BIBdecl

\bibitem{cha2006case}
M.~Cha, G.~Choudhury, J.~Yates, A.~Shaikh, and S.~Moon, ``Case study: Resilient
  backbone design for iptv services,'' in \emph{Proceedings of the IPTV
  Workshop, International World Wide Web Conference}, 2006.

\bibitem{TwitchViewers2015}
``2015 retrospective - twitch,'' \url{https://www.twitch.tv/year/2015},
  accessed: 24-11-2016.

\bibitem{diot2000deployment}
C.~Diot, B.~N. Levine, B.~Lyles, H.~Kassem, and D.~Balensiefen, ``Deployment
  issues for the ip multicast service and architecture,'' \emph{IEEE network},
  vol.~14, no.~1, pp. 78--88, 2000.

\bibitem{li2013scaling}
X.~Li and M.~J. Freedman, ``Scaling ip multicast on datacenter topologies,'' in
  \emph{Proceedings of the ninth ACM conference on Emerging networking
  experiments and technologies}.\hskip 1em plus 0.5em minus 0.4em\relax ACM,
  2013, pp. 61--72.

\bibitem{iyer2014avalanche}
A.~Iyer, P.~Kumar, and V.~Mann, ``Avalanche: Data center multicast using
  software defined networking,'' in \emph{2014 Sixth International Conference
  on Communication Systems and Networks (COMSNETS)}.\hskip 1em plus 0.5em minus
  0.4em\relax IEEE, 2014, pp. 1--8.

\bibitem{bondan2013multiflow}
L.~Bondan, L.~F. M{\"u}ller, and M.~Kist, ``Multiflow: Multicast clean-slate
  with anticipated route calculation on openflow programmable networks,''
  \emph{Journal of Applied Computing Research}, vol.~2, no.~2, pp. 68--74,
  2013.

\bibitem{arefin2013opensession}
A.~Arefin, R.~Rivas, R.~Tabassum, and K.~Nahrstedt, ``Opensession: Sdn-based
  cross-layer multi-stream management protocol for 3d teleimmersion,'' in
  \emph{2013 21st IEEE International Conference on Network Protocols
  (ICNP)}.\hskip 1em plus 0.5em minus 0.4em\relax IEEE, 2013, pp. 1--10.

\bibitem{zhao2014software}
M.~Zhao, B.~Jia, M.~Wu, H.~Yu, and Y.~Xu, ``Software defined network-enabled
  multicast for multi-party video conferencing systems,'' in \emph{2014 IEEE
  International Conference on Communications (ICC)}.\hskip 1em plus 0.5em minus
  0.4em\relax IEEE, 2014, pp. 1729--1735.

\bibitem{noghani2014streaming}
K.~A. Noghani and M.~O. Sunay, ``Streaming multicast video over
  software-defined networks,'' in \emph{2014 IEEE 11th International Conference
  on Mobile Ad Hoc and Sensor Systems}.\hskip 1em plus 0.5em minus 0.4em\relax
  IEEE, 2014, pp. 551--556.

\bibitem{egilmez2012openqos}
H.~E. Egilmez, S.~T. Dane, K.~T. Bagci, and A.~M. Tekalp, ``Openqos: An
  openflow controller design for multimedia delivery with end-to-end quality of
  service over software-defined networks,'' in \emph{Signal \& Information
  Processing Association Annual Summit and Conference (APSIPA ASC), 2012
  Asia-Pacific}.\hskip 1em plus 0.5em minus 0.4em\relax IEEE, 2012, pp. 1--8.

\bibitem{van2014fast}
N.~L.~M. van Adrichem, B.~J. van Asten, and F.~A. Kuipers, ``Fast recovery in
  software-defined networks,'' in \emph{2014 Third European Workshop on
  Software Defined Networks}.\hskip 1em plus 0.5em minus 0.4em\relax IEEE,
  2014, pp. 61--66.

\bibitem{Niels2}
N.~L.~M. van Adrichem, F.~Iqbal, and F.~A. Kuipers, ``Backup rules in
  software-defined networks,'' in \emph{IEEE Conference on Network Function
  Virtualization and Software Defined Networks}, 2016.

\bibitem{OpenFlowSite}
``Openflow - open networking foundation,''
  \url{https://www.opennetworking.org/sdn-resources/openflow}.

\bibitem{mckeown2008openflow}
N.~McKeown, T.~Anderson, H.~Balakrishnan, G.~Parulkar, L.~Peterson, J.~Rexford,
  S.~Shenker, and J.~Turner, ``Openflow: enabling innovation in campus
  networks,'' \emph{ACM SIGCOMM Computer Communication Review}, vol.~38, no.~2,
  pp. 69--74, 2008.

\bibitem{recodis}
C.~Mas~Machuca, S.~Secci, P.~Vizarreta, F.~Kuipers, A.~Gouglidis, D.~Hutchison,
  S.~Jouet, D.~Pezaros, A.~Elmokashfi, P.~Heegaard, and S.~Ristov,
  ``Technology-related disasters: A survey towards disaster-resilient software
  defined networks,'' in \emph{Proceedings of the 8th International Workshop on
  Reliable Networks Design and Modeling (RNDM 2016)}, 2016.

\bibitem{kuiperssurvey}
F.~A. Kuipers, ``An overview of algorithms for network survivability,''
  \emph{ISRN Communications and Networking}, vol. 2012, 2012.

\bibitem{staessens2011software}
D.~Staessens, S.~Sharma, D.~Colle, M.~Pickavet, and P.~Demeester, ``Software
  defined networking: Meeting carrier grade requirements,'' in \emph{Local \&
  Metropolitan Area Networks (LANMAN), 2011 18th IEEE Workshop on}.\hskip 1em
  plus 0.5em minus 0.4em\relax IEEE, 2011, pp. 1--6.

\bibitem{sharma2013openflow}
S.~Sharma, D.~Staessens, D.~Colle, M.~Pickavet, and P.~Demeester, ``Openflow:
  Meeting carrier-grade recovery requirements,'' \emph{Computer
  Communications}, vol.~36, no.~6, pp. 656--665, 2013.

\bibitem{kuzniar2013automatic}
M.~Ku{\'z}niar, P.~Pere{\v{s}}{\'\i}ni, N.~Vasi{\'c}, M.~Canini, and
  D.~Kosti{\'c}, ``Automatic failure recovery for software-defined networks,''
  in \emph{Proceedings of the second ACM SIGCOMM workshop on Hot topics in
  software defined networking}.\hskip 1em plus 0.5em minus 0.4em\relax ACM,
  2013, pp. 159--160.

\bibitem{lee2014path}
S.~S. Lee, K.-Y. Li, K.-Y. Chan, G.-H. Lai, and Y.-C. Chung, ``Path layout
  planning and software based fast failure detection in survivable openflow
  networks,'' in \emph{Design of Reliable Communication Networks (DRCN), 2014
  10th International Conference on the}.\hskip 1em plus 0.5em minus 0.4em\relax
  IEEE, 2014, pp. 1--8.

\bibitem{kempf2012scalable}
J.~Kempf, E.~Bellagamba, A.~Kern, D.~Jocha, A.~Tak{\'a}cs, and
  P.~Sk{\"o}ldstr{\"o}m, ``Scalable fault management for openflow,'' in
  \emph{2012 IEEE International Conference on Communications (ICC)}.\hskip 1em
  plus 0.5em minus 0.4em\relax IEEE, 2012, pp. 6606--6610.

\bibitem{sgambelluri2013openflow}
A.~Sgambelluri, A.~Giorgetti, F.~Cugini, F.~Paolucci, and P.~Castoldi,
  ``Openflow-based segment protection in ethernet networks,'' \emph{Journal of
  Optical Communications and Networking}, vol.~5, no.~9, pp. 1066--1075, 2013.

\bibitem{capone2015detour}
A.~Capone, C.~Cascone, A.~Q. Nguyen, and B.~Sanso, ``Detour planning for fast
  and reliable failure recovery in sdn with openstate,'' in \emph{Design of
  Reliable Communication Networks (DRCN), 2015 11th International Conference on
  the}.\hskip 1em plus 0.5em minus 0.4em\relax IEEE, 2015, pp. 25--32.

\bibitem{reitblatt2013fattire}
M.~Reitblatt, M.~Canini, A.~Guha, and N.~Foster, ``Fattire: Declarative fault
  tolerance for software-defined networks,'' in \emph{Proceedings of the second
  ACM SIGCOMM workshop on Hot topics in software defined networking}.\hskip 1em
  plus 0.5em minus 0.4em\relax ACM, 2013, pp. 109--114.

\bibitem{kitsuwan2015independent}
N.~Kitsuwan, S.~McGettrick, F.~Slyne, D.~B. Payne, and M.~Ruffini,
  ``Independent transient plane design for protection in openflow-based
  networks,'' \emph{Journal of Optical Communications and Networking}, vol.~7,
  no.~4, pp. 264--275, 2015.

\bibitem{borokhovich2014provable}
M.~Borokhovich, L.~Schiff, and S.~Schmid, ``Provable data plane connectivity
  with local fast failover: Introducing openflow graph algorithms,'' in
  \emph{Proceedings of the third workshop on Hot topics in software defined
  networking}.\hskip 1em plus 0.5em minus 0.4em\relax ACM, 2014, pp. 121--126.

\bibitem{gyllstrom2014recovery}
D.~Gyllstrom, N.~Braga, and J.~Kurose, ``Recovery from link failures in a smart
  grid communication network using openflow,'' in \emph{Smart Grid
  Communications (SmartGridComm), 2014 IEEE International Conference on}.\hskip
  1em plus 0.5em minus 0.4em\relax IEEE, 2014, pp. 254--259.

\bibitem{kotani2012design}
D.~Kotani, K.~Suzuki, and H.~Shimonishi, ``A design and implementation of
  openflow controller handling ip multicast with fast tree switching,'' in
  \emph{Applications and the Internet (SAINT), 2012 IEEE/IPSJ 12th
  International Symposium on}.\hskip 1em plus 0.5em minus 0.4em\relax IEEE,
  2012, pp. 60--67.

\bibitem{DaisukeKotani2016}
------, ``A multicast tree management method supporting fast failure recovery
  and dynamic group membership changes in openflow networks,'' \emph{Journal of
  Information Processing}, vol.~24, no.~2, pp. 395--406, 2016.

\bibitem{subtree}
V.~R. Raja, C.-H. Lung, A.~Pandey, G.~ming Wei, and A.~Srinivasan, ``A
  subtree-based approach to failure detection and protection for multicast in
  sdn,'' \emph{Frontiers of Information Technology \& Electronic Engineering},
  vol.~17, no.~7, pp. 682--700, 2016.

\bibitem{pfeiffenberger2015reliable}
T.~Pfeiffenberger, J.~L. Du, P.~B. Arruda, and A.~Anzaloni, ``Reliable and
  flexible communications for power systems: fault-tolerant multicast with
  sdn/openflow,'' in \emph{2015 7th International Conference on New
  Technologies, Mobility and Security (NTMS)}.\hskip 1em plus 0.5em minus
  0.4em\relax IEEE, 2015, pp. 1--6.

\bibitem{doi:10.1137/0404033}
\BIBentryALTinterwordspacing
M.~Imase and B.~M. Waxman, ``Dynamic steiner tree problem,'' \emph{SIAM Journal
  on Discrete Mathematics}, vol.~4, no.~3, pp. 369--384, 1991. [Online].
  Available: \url{http://dx.doi.org/10.1137/0404033}
\BIBentrySTDinterwordspacing

\bibitem{JorikApp}
``Openflow application providing {$F$}-link fault tolerance for multicast
  streams,'' \url{https://github.com/TUDelftNAS/SDN-ResilientMulticast}.

\bibitem{Ryu}
``Ryu,'' \url{https://osrg.github.io/ryu/}.

\bibitem{NetworkX}
``Networkx,'' \url{https://networkx.github.io/}.

\bibitem{GEANTTopology}
``G\'eant topology map,'' october 2015.

\bibitem{OpenVSwitch}
``Open vswitch,'' \url{http://openvswitch.org/}.

\bibitem{huang2013high}
D.~Y. Huang, K.~Yocum, and A.~C. Snoeren, ``High-fidelity switch models for
  software-defined network emulation,'' in \emph{Proceedings of the second ACM
  SIGCOMM workshop on Hot topics in software defined networking}.\hskip 1em
  plus 0.5em minus 0.4em\relax ACM, 2013, pp. 43--48.

\bibitem{HPGuide}
``Hp switch software openflow v1.3 administrator guide k/ka/kb/wb 15.18,''
  august 2015.

\end{thebibliography}
\end{document}